\newcommand{\be}{\begin{equation}}\newcommand{\ee}{\end{equation}}
\newcommand{\bea}{\begin{eqnarray}}\newcommand{\eea}{\end{eqnarray}}

\documentstyle[12pt]{article}
\def\theequation{\arabic{section}.\arabic{equation}}
\begin{document}
\thispagestyle{empty}
\begin{flushright}BONN-TH-94-02 \\
hep-th/xxxxxx \\
April 1994
 \end{flushright}
\vskip 1.0truecm
\begin{center}{\bf\Large
Sigma models in (4,4) harmonic superspace}\end{center}
 \vskip 1.0truecm
\centerline{\bf E. Ivanov${}^{(a,b)}$ and A. Sutulin${}^{(b)}$}
\vskip 1.0truecm
\centerline{${}^{(a)}$ \it Physicalisches Institut, Universit\"at
Bonn}
\centerline{\it Nussallee 12, 5300 Bonn 1, Germany}
\vskip5mm
\centerline{${}^{(b)}$\it Bogoliubov Laboratory of
Theoretical Physics, JINR,}
\centerline{\it 141 980 Dubna (Moscow region),
Russian Federation}
\vskip 1.0truecm  \nopagebreak

\begin{abstract}
We define basics of $(4,4)\;\; 2D$ harmonic superspace with two
independent
sets of $SU(2)$ harmonic variables and apply it
to construct new superfield actions of $(4,4)$ supersymmetric
two-dimensional sigma models with torsion and mutually commuting
left and
right complex
structures, as well as of their
massive deformations. We show that the generic off-shell
sigma model action is the general action of
constrained analytic superfields $q^{(1,1)}$ representing twisted
$N=4$ multiplets in $(4,4)$ harmonic superspace. The massive term of
$q^{(1,1)}$ is shown to be unique; it generates a
scalar potential the form of which is determined by the metric
on the target bosonic manifold. We discuss in detail $(4,4)$
supersymmetric group manifold $SU(2)\times U(1)$ WZNW sigma model
and its Liouville deformation. A deep analogy of the relevant
superconformally invariant analytic superfield action to that of the
improved tensor $N=2\;\;4D$ multiplet is found. We define
$(4,4)$ duality transformation
and find new off-shell dual representations of the previously
constructed
actions via {\it unconstrained} analytic $(4,4)$ superfields.
The main
peculiarities of the $(4,4)$ duality transformation are:
(i) it preserves
manifest $(4,4)$ supersymmetry; (ii) dual actions reveal a
gauge invariance needed for the on-shell equivalence to the original
description;
(iii) in the actions dual to the massive ones $2D$
supersymmetry is modified off shell by $SU(2)$ tensor central charges.
The dual representation
suggests some hints of how to describe $(4,4)$
models with non-commuting complex structures in the harmonic superspace.

\end{abstract}
\newpage\setcounter{page}1

\section{Introduction}
Two-dimensional $(4,4)$ supersymmetric sigma models with torsion [1 - 5]
have a number of interesting applications, e.g. in the string theory
(see, e.g., \cite{Callan}) and the $2D$ black hole business
\cite{{RSS},{RASS}}. An important subclass of these models
is provided by those on group manifolds, that is,
by $(4,4)$ supersymmetric WZNW sigma models
\cite{{IK2},{IKL},{Belg1},{Belg2}}.
A superconformally invariant deformation of the simplest model of that
kind, with $SU(2)\times U(1)$ as the bosonic target, is
the $N=4\;\; SU(2)$ super
Liouville theory \cite{{IK2},{IKL},{GI}}. It proved to be the first
example of integrable
$N=4 \;\; 2D$ system and it is expected to play a crucial role in
$N=4 \;\;2D$ induced supergravity \cite{{Rostand},{Troost}}.
Presumably, $N=4$ superextensions of
other integrable bosonic systems, such as the Toda and
sine-Gordon ones,
can also be obtained as some deformations of the appropriate
$(4,4)$ WZNW sigma models. These theories could encode a rich set
of invariances and, by analogy with $N=4$ super Liouville model,
be related to $N=4$ superextensions of induced $W$ gravities.

To explore all these exciting issues in a full generality and to
clearly understand their underlying geometric aspects, one needs a
convenient off-shell superfield description of the
$(4,4)$ sigma models in question.

Until now such superfield formulations were given in the $(2,2)$
and projective $(4,4)$ superspaces
\cite{{GHR},{RSS},{RASS},{BLR},{RLI}}, as well as in the conventional
$(4,4)$ superspace \cite{GI}. Basically, the simplest case of
sigma models with mutually commuting left and right complex structures
was treated.
As for the more general class of models with non-commuting structures,
at present only
some proposals exist how to describe them in the superfield
terms \cite{{BLR},{RLI}}. On the other hand, an adequate
framework for theories with extended supersymmetry ($N \geq 2$ in $4D$,
$N \geq 3$ in $2D$) is provided by the harmonic superspace approach
\cite{GIKOS}.
Though up to now its main applications concerned supersymmetric
theories in four dimensions, it is quite natural to expect that
this approach
is applicable with equal efficiency to theories with two-dimensional
extended supersymmetry, including the aforementioned
$(4,4)$ sigma models.
We believe that the most appropriate arena to handle the problems
mentioned in the beginning is just the $(4,4)$ harmonic superspace.

In the present paper we define basic elements of the $(4,4)$ harmonic
superspace calculus in two dimensions and apply it to construct new
off-shell superfield
formulations of $(4,4)\;\;2D$ sigma models with torsion as well as
of their massive deformations. We limit our
consideration to the simplest case of mutually
commuting left and right complex structures, the more general class
of models with non-commuting structures will be attacked in
further publications. Here (in the end of Sect.6) we make only some
comments on possible ways of describing these models in the harmonic
superspace.

We start in Sect.2 by giving generalities of $(4,4)$ harmonic
superspace and its analytic subspace which has twice as few
Grassmann coordinates. The most characteristic feature of this formalism
is the presence of
two sets of
harmonic variables which are associated with the $SU(2)$ automorphism
groups of two $2D$ light-cone copies of $N=4$ supersymmetry.
Among other things, we show
that there exist two
different $N=4 \;\; SU(2)$ superconformal groups preserving harmonic
$(4,4)$ analyticity, their closure being
$N=4 \;\;SO(4)\times U(1)$ (``large'') superconformal group.

In  Sect.3 the general form
of the $(4,4)$ sigma model action with two commuting sets of complex
structures is given as an integral over an analytic subspace of
$(4,4)$ harmonic superspace. This action is the general action of
the analytic superfield $q^{(1,1)}$ which has a constrained harmonic
dependence and represents the twisted $(4,4)$ supermultiplet
\cite{{GHR},{IK2},{IKL}}
in the harmonic superspace. The relevant
component action
is shown to be completely specified by the metric on the physical
bosons manifold.
The general action is always invariant under one of
two $N=4 \;\; SU(2)$ superconformal groups defined in Sect.2, with the
$SU(2)$ Kac-Moody subgroup acting only on fermions.

In Sect.4 we show that the requirement of invariance under another
superconformal group the
$SU(2)$ Kac-Moody subgroup of which is realized both on the bosonic
and fermionic
fields, uniquely fixes the action to be that of $(4,4)$
superextension of the group manifold $SU(2)\times U(1)$ WZNW sigma
model. The action has an unexpectedly
simple form and bears a deep analogy with the $N=2\;\;4D$ harmonic
superspace
action of the improved tensor $N=2$ multiplet \cite{GIO1}. We discuss
some unusual properties of the action constructed and explain
in detail the process of descending to components.

Sect.5 is devoted to constructing massive deformations of the
superfield actions presented in the previous Sections. We
find that, without allowing for central charges, the massive term of
superfields $q^{(1,1)}$ is defined
uniquely. It preserves one of the superconformal groups mentioned above,
namely the one with $SU(2)$ acting both on bosons and fermions.
Being added
to the $SU(2)\times U(1)$ action, it produces the $N=4\;\; SU(2)$
WZNW - Liouville model of ref. \cite{{IK2},{IKL},{GI}}. In general
it adds to the
physical component action the potential terms strictly specified
by the form of the $(4,4)$ sigma model action one started with.

In Sect.6 we define the $(4,4)$ duality transformation:
insert the harmonic constraints the
superfields $q^{(1,1)}$ satisfy into the action and then
find a dual form of the action in terms of the relevant unconstrained
analytic Lagrange multiplier superfields. This
duality transformation, in contrast to the one used in
ref. \cite{{RSS},{RASS}}, manifestly preserves $(4,4)$ supersymmetry.
The crucially
new feature of the dual action is the presence of an infinite
number of
auxiliary fields. While in the original action the correct
physical field
content is ensured by the harmonic constraints, in the dual action
this is
achieved due to an appropriate gauge
invariance. We explicitly give the dual
form of the $SU(2)\times U(1)$ action with and without Liouville term.
An interesting peculiarity of the dual form of massive actions
is the
spontaneous generation of ``semi-central'' charges breaking
the commutativity
of the left and right light-cone $(4,4)$ supertranslations and
possessing
nontrivial tensor properties with respect to the automorphism $SU(2)$'s.

\setcounter{equation}{0}
\section{Generalities of (4,4) harmonic superspace}

{\bf 2.1 Conventional (4,4) superspace.} We start with the basic
relations of the algebra of $(4,4)\; \;2D$
covariant spinor derivatives\footnote{We use the standard conventions:
$SU(2)$ doublet indices are raised and lowered with the help of totally
antisymmetric tensors $\epsilon_{ik},\; \epsilon_{ab}, \;\epsilon^{ik},
\epsilon^{ab},\; (\epsilon_{12} = - \epsilon^{12} = 1)$, etc.}
\be
\{ D_{+\;i}, \overline{D}^{j}_{+} \} = 2i\delta^j_i \partial_{++},\;\;
\{ D_{-\;a}, \overline{D}^{b}_{-} \} = 2i\delta^b_a \partial_{--}, \;\;
\{ D_{+\;i}, D_{-\;a} \} = \{ D_{+\;i}, \overline{D}^{a}_{-} \} = 0\;.
\label{comcovd}
\ee
Here
\bea
D_{+\;i} &=& \frac{\partial}{\partial \theta^{+\;i}} +
i \bar{\theta}^{+}_i\partial_{++},\;\; \overline{D}^{\;i}_{+} \;=\; -
\frac{\partial}{\partial \bar \theta^{+}_{i}} -
i \theta^{+\;i}\partial_{++}
\nonumber \\
D_{-\;a} &=& \frac{\partial}{\partial \theta^{-\;a}} +
i \bar{\theta}^{-}_a
\partial_{--},\;\; \overline{D}^{\;a}_{-} \;=\; -
\frac{\partial}{\partial \bar{\theta}^{-}_{a}} -
i \theta^{-\;a}\partial_{--}
\nonumber \\
\partial_{\pm \pm} &=& \frac{\partial}{\partial x^{\pm \pm}}\;.
\label{defcovd}
\eea
The $2D$ Lorentz indices ``$+, \;-$'' mark quantities related to
the left and right light-cone sectors of the $(4,4)\;\; 2D$ superspace,
\be
{\bf S}^{(1,1|4,4)} \equiv  {\bf L}^{(1|4)} \otimes {\bf R}^{(1|4)} \;,
\ee
\be
{\bf L}^{(1|4)} = \{ x^{++}, \theta^{+ \;i},
\bar \theta^{+}_{j} \} \equiv \{ Z_{L} \} \;,
\;\; {\bf R}^{(1|4)} = \{ x^{--}, \theta^{-\;a}, \bar \theta^{-}_{b} \}
\equiv  \{ Z_{R} \}\;,
\ee
the isodoublet indices $i,j = 1,2$ and $a,b = 1,2$ are associated
with two
independent automorphism $SU(2)$ groups acting in the left
and right sectors.

The left $(4,4)\;\; 2D$ Poincar\'e supertranslations are realized
on the coordinates of ${\bf L}^{(1|4)}$ by
\be
x^{++\;'} = x^{++} + i(\theta^{+\;i}\bar \epsilon^{+}_i -
\epsilon^{+\;i}\bar \theta^{+}_{i})\;,\;\;\; \theta^{+\;i\;'} =
\theta^{+\;i} + \epsilon^{+\;i}\;, \bar{\theta}^{+\;'}_{i} =
\bar{\theta}^{+}_i + \bar{\epsilon}^{+}_i\;,
\label{susy}
\ee
where $\epsilon^{+\;i}$ is the related constant parameter. The
realization of the right supertranslations on the coordinates of
${\bf R}^{(1|4)}$ has the same form, up to
the replacements $+\rightarrow -\;, i\rightarrow a$.

For our further purposes it will be of crucial importance that
$(4,4)\;\;2D$ supersymmetry
possesses two commuting automorphism groups $SU(2)_L$ and
$SU(2)_R$ acting, respectively, on the doublet indices $i,j$
and $a,b$.
Note that, besides these explicit $SU(2)$'s, the relations
(2.1) reveal
covariance with respect to another two $SU(2)$ groups which
rotate spinor
quantities through their conjugates and commute with each
other and with the explicit $SU(2)$'s. All these $SU(2)$
automorphisms
(forming the group $SO(4)_L \times SO(4)_R$) become manifest in the
quartet notation (we suppress the light-cone indices)
\bea
(\theta^i, \bar{\theta}^i) &\equiv & \theta^{i\underline{k}}\;, \;\;
\;\;(D_i, \overline{D}_i) \;\equiv \; D_{i\underline{k}}\;,\;\;
\underline{k} = 1,2 \nonumber \\
(\theta^a, \bar{\theta}^a) &\equiv & \theta^{a\underline{b}}\;,
 \;\;
\;\;(D_a, \overline{D}_a) \;\equiv \; D_{a\underline{b}}\;,\;\;
\underline{b} = 1,2 \nonumber \\
(\theta^{i\underline{l}})^{\dagger} &=& \epsilon_{ik}
\epsilon_{\underline{l}
\underline{m}}\theta^{k\underline{m}}\;,
\;\;(D_{i\underline{k}})^{\dagger} \;=\;
- \epsilon^{ik} \epsilon^{\underline{l}
\underline{m}} D_{k\underline{m}}\;,\; etc.
\label{quartet}
\eea
For instance, in this notation the algebra of
spinor derivatives belonging to the left world can be summarized as
the single relation
\be
\{ D_{+\;i\underline{k}}, D_{+\;j\underline{l}} \} = -2i
\epsilon_{ij} \epsilon_{\underline{k}\underline{l}} \partial_{++}\;.
\ee

Actually, the harmonic extension of $(4,4)$ superspace we
will deal with in the present paper uses the harmonic variables
on the automorphism
groups $SU(2)_L$ and $SU(2)_R$ which are explicit in the doublet
notation (2.1), (2.2). For our purposes it will be of no need to
harmonize two other $SU(2)$'s, they will be
treated as additional automorphism groups of the $(4,4)$ harmonic
superspace and an analytic
subspace of the latter.

\vspace{0.4cm}
\noindent{\bf 2.2 Harmonic (4,4) superspace.} In constructing a
harmonic extension of the $(4,4)\;\;2D$
superspace ${\bf S}^{(1,1|4,4)}$ we will closely follow the lines of
ref.
\cite{GIKOS}. The main new feature of the present case is the
possibility
to introduce two independent sets of harmonic variables associated
with the mutually commuting automorphism groups $SU(2)_L$, $SU(2)_R$.

Thus we define the harmonic $(4,4)$ superspace
${\bf HS}^{(1+2,1+2|4,4)}$ as
\be
{\bf HS}^{(1+2,1+2|4,4)} = {\bf HL}^{(1+2|4)} \otimes
{\bf HR}^{(1+2|4)} \;,
\ee
\be
{\bf HL}^{(1+2|4)} = {\bf L}^{(1|4)} \otimes {\bf S}^2_L
\equiv \{ Z_L, u^{(\pm 1)\;i} \} \;,\;\;
{\bf HR}^{(1+2|4)} = {\bf R}^{(1|4)} \otimes {\bf S}^2_R \equiv
\{ Z_R, v^{(\pm 1)\;a} \}\;.
\label{defhs}
\ee
Here the harmonic coordinates $u^{(1)\;i}, u^{(-1)\;i}$ and
$v^{(1)\;a},
v^{(-1)\;a}$ parametrize two-dimensional spheres ${\bf S}^2_L$,
${\bf S}^2_R$:
\be
{\bf S}^2_L \sim SU(2)_L /U(1)_L =
\{ u^{(1)\;i},\; u^{(-1)\;j} \}\;, \;\;
{\bf S}^2_R \sim SU(2)_R /U(1)_R = \{ v^{(1)\;a},\; v^{(-1)\;b} \}
\ee
\be
u^{(1)\;i} u^{(-1)}_i =1\;,\;\;\; v^{(1)\;a} v^{(-1)}_a = 1\;.
\label{unit}
\ee
Actually, each set of harmonics brings just two independent parameters
in agreement with the dimension of the cosets
${\bf S}^2_L$, ${\bf S}^2_R$.
One of three parameters remaining after employing the
unitarity conditions
(\ref{unit}) does not
contribute because we require the strict preservation of
the relevant $U(1)$ charge (such a
requirement is standard for the harmonic superspace
approach \cite{GIKOS}).

One may define, for each set of the harmonic variables,
three derivatives
compatible with the conditions (\ref{unit})
\bea
D^{(\pm 2,0)} &=& u^{(\pm1)\;i}\frac{\partial}
{\partial u^{(\mp1)\;i}}\;,\;\;
D_u^{(0,0)} \;=\; u^{(1)\;i}\frac{\partial}{\partial u^{(1)\;i}} -
u^{(-1)\;i}\frac{\partial}{\partial u^{(-1)\;i}}\; \nonumber \\
D^{(0,\pm 2)} &=& v^{(\pm 1)\;a}\frac{\partial}
{\partial v^{(\mp 1)\;a}}\;,
\; D_v^{(0,0)} \;=\; v^{(1)\;a}\frac{\partial}{\partial v^{(1)\;a}} -
v^{(-1)\;a}\frac{\partial}{\partial v^{(-1)\;a}}\;.
\label{harmder}
\eea
They form two commuting algebras $su(2)$
\be
[ D^{(2,0)}, D^{(-2,0)}] \;=\; D^{(0,0)}_u\;,\;\;
[ D^{(0,0)}_u, D^{(\pm 2, 0)} ]
\;=\; \pm 2 D^{(\pm 2, 0)}\;, \nonumber
\ee
\be
[ D^{(0,2)}, D^{(0,-2)}] \;=\; D^{(0,0)}_v \;,\;\;
[ D^{(0,0)}_v, D^{(0, \pm 2)} ]
\;=\; \pm 2 D^{(0, \pm 2)}\;.
\label{alghder}
\ee
In general, the superfields given on the superspace (\ref{defhs}),
harmonic $(4,4)$ superfields, are characterized by two
$U(1)$ charges which
are eigenvalues of the operators $D^{(0,0)}_u\;, D^{(0,0)}_v$
\be
\Phi^{(q,p)} = \Phi^{(q,p)} ( Z_L, Z_R, u, v )\;,\;\;
D^{(0,0)}_u \Phi^{(q,p)} = q \Phi^{(q,p)}\;,\;\;
D^{(0,0)}_v \Phi^{(q,p)} =
p \Phi^{(q,p)}\;.
\ee
These superfields in general contain infinite numbers of components
coming from two independent harmonic expansions on the two-spheres
${\bf S}^2_L$ and ${\bf S}^2_R$, i.e. with respect to the
harmonics $u$ and $v$. For instance
\be
\Phi^{(1,1)}(Z_L,Z_R, u,v) = \Phi^{ia}(Z_L,Z_R) u^{(1)}_i
v^{(1)}_a + ...
\ee

\vspace{0.4cm}
\noindent{\bf 2.3 (4,4) harmonic analyticity.} Let us pass to
another
(analytic) basis in the left and right harmonic superspaces
(\ref{defhs})
\bea
{\bf HL}^{(1+2|4)} &=& \{ z^{++}, \theta^{+\;(\pm1,0)}, \bar
\theta^{+\;(\pm1,0)}, u^{(\pm 1)\;i} \}\;, \nonumber \\
{\bf HR}^{(1+2|4)} &=& \{ z^{--}, \theta^{-\;(0, \pm1)},
\theta^{-\;(0, \pm1)}, v^{(\pm 1)\;a} \}\;,
\eea
where
\bea
z^{\pm \pm} &=& x^{\pm \pm} + i (\theta^{\pm \;(1,0)} \bar
\theta^{\pm \;(-1,0)} +
\theta^{\pm \;(-1,0)} \bar \theta^{\pm \;(1,0)}) \nonumber \\
\theta^{+ \;(\pm 1,0)} &=& \theta^{+ \;i} u_{i}^{(\pm 1)}\;,\;
\bar \theta^{+ \;(\pm 1,0)} \;=\; \bar \theta^{+ \;i}
u_{i}^{(\pm 1)}\;,\nonumber \\
\theta^{- \;(0,\pm 1)} &=& \theta^{- \;a} v_{a}^{(\pm 1)}\;,\;
\bar \theta^{- \;(0,\pm 1)} \;=\; \bar \theta^{- \;a} v_{a}^{(\pm 1)}\;.
\label{anbas}
\eea
Henceforth, for brevity, we will omit the light-cone Lorentz indices
of spinor
coordinates in the analytic basis.

It is easy to check that the coordinate sets
\bea
{\bf AL}^{(1+2|2)} &=& \{ z^{++}, \theta^{(1,0)}, \bar \theta^{(1,0)},
u^{(\pm 1)\;i} \} \;\equiv\; \{ \zeta_L, u \}\;, \nonumber \\
{\bf AR}^{(1+2|2)} &=& \{ z^{--}, \theta^{(0,1)}, \bar \theta^{(0,1)},
v^{(\pm 1)\;a} \} \;\equiv\; \{ \zeta_R, v \}
\label{anlr}
\eea
are closed under the $(4,4)$ supersymmetry transformations (\ref{susy})
and so form invariant analytic subspaces in the above
harmonic superspaces.
Their product is the analytic harmonic $(4,4)$ superspace \footnote{To
avoid a confusion, we point out that the symbol $(4,4)$ indicates the
numbers of left and right supersymmetries, but not the Grassmann
dimension of superspaces where these supersymmetries are realized.}
\be
{\bf AS}^{(1+2,1+2|2,2)} = {\bf AL}^{(1+2|2)} \otimes
{\bf AR}^{(1+2|2)} =
\{ \zeta_L, \zeta_R, u, v \} \equiv \{ \zeta, u,v \}\;.
\label{anharm}
\ee
The existence of this analytic subspace matches with the form of
covariant
spinor derivatives in the analytic basis
\bea
D^{(\pm 1,0)} &\equiv& D^i u^{(\pm 1)}_i\;,\;\;
\;\;\;\;\;\overline{D}^{(\pm 1, 0)} \;\equiv\; \overline{D}^i
u^{(\pm 1)}_i\;,
\nonumber \\
D^{(1,0)} &=& -\frac{\partial}{\partial \theta^{(-1,0)}}\;,\;
\;\;\;\;\;\;\overline{D}^{(1,0)} \;=\; -\frac{\partial}{\partial \bar
\theta^{(-1,0)}}
\nonumber \\
D^{(-1,0)} &=& \frac{\partial}{\partial \theta^{(1,0)}} +2i\;
\bar \theta^{(-1,0)}
\partial_{++}\;, \;\;
\overline{D}^{(-1,0)} \;=\; \frac{\partial}{\partial \bar
\theta^{(1,0)}} - 2i\; \theta^{(-1,0)}
\partial_{++}
\label{covdan}
\eea
($D^{(0,\pm 1)} \equiv D^a v^{(\pm 1)}_a$ and $\overline{D}^{(0,\pm 1)}
\equiv \overline{D}^a v^{(\pm 1)}_a$ are given by analogous formulas).
Harmonic superfields obeying the $(4,4)$ harmonic Grassmann analyticity
conditions
\be
D^{(1,0)} \Psi^{(q,p)} = \overline{D}^{(1,0)} \Psi^{(q,p)} =
D^{(0,1)} \Psi^{(q,p)} = \overline{D}^{(0,1)} \Psi^{(q,p)} = 0
\label{ancon}
\ee
are called analytic $(4,4)$ superfields. In the basis
(\ref{anbas}) the
spinor derivatives entering (\ref{ancon}) are reduced to the
partial ones,
so the conditions (\ref{ancon}) mean that the analytic
superfields
do not depend on the coordinates $\theta^{(-1,0)},
\bar \theta^{(-1,0)},
\theta^{(0,-1)}, \bar \theta^{(0,-1)}$ in this basis
\be
\Psi^{(q,p)} = \Psi^{(q,p)} (\zeta_L, \zeta_R, u, v)\;.
\ee

In the sequel we will need the expressions for the harmonic
derivatives
$D^{(2,0)}, D^{(0,2)}$ in the basis (\ref{anbas}) and in the
realization on
analytic superfields
\be
D^{(2,0)} = \partial^{(2,0)} +2 i \;\theta^{(1,0)} \bar \theta^{(1,0)}
\partial_{++}\;,\;\;\;
D^{(0,2)} = \partial^{(0,2)} +2 i\; \theta^{(0,1)} \bar \theta^{(0,1)}
\partial_{--}\;,
\label{anhder}
\ee
where partial harmonic derivatives are given by the expressions
(\ref{harmder}). We see that these harmonic derivatives (as well as
$D^{(0,0)}_u$, $D_v^{(0,0)}$) preserve $(4,4)$ analyticity:
the result of their action on an analytic superfield is
again an analytic superfield. For completeness, we also present
the analytic superspace form of the $U(1)$ charge operators
\bea
D^{(0,0)}_u &=& \partial^{(0,0)}_u + \theta^{(1,0)}\frac{\partial}
{\partial \theta^{(1,0)}} + \bar \theta^{(1,0)} \frac{\partial}
{\partial \bar \theta^{(1,0)}}\;, \nonumber \\
D^{(0,0)}_v &=& \partial^{(0,0)}_v + \theta^{(0,1)}\frac{\partial}
{\partial \theta^{(0,1)}} + \bar \theta^{(0,1)} \frac{\partial}
{\partial \bar \theta^{(0,1)}}\;.
\eea

Finally, we note that, like in the $N=2\;\;4D$ case \cite{GIKOS},
the analytic
subspace ${\bf AL}^{(1+2|2)}$ (${\bf AR}^{(1+2|2)}$) is real
with respect to the
generalized involution ``$\sim $'' which is the product of
ordinary complex conjugation
and an antipodal map of the sphere ${\bf S}^2_L$ (${\bf S}^2_R$)
\be
\widetilde{(\theta^{(1,0)})} = - \bar \theta^{(1,0)}\;,\;\;
\widetilde{(\bar \theta^{(1,0)})} = \theta^{(1,0)}\;,\;\;
\widetilde{(u^{(\pm 1)\;i})} = - u^{(\pm 1)}_i\;,
\ee
(and similarly for $\theta^{(0,1)}, \bar \theta^{(0,1)},
v^{(\pm 1)}_a$).
The analytic superfields $\Psi^{(p,q)} (\zeta_L, \zeta_R, u, v)$
can be chosen real with respect to this involution, provided
$|p+q| = 2n$
\be
\widetilde{(\Psi^{(p,q)})} = \Psi^{(p,q)}\;, |p+q| = 2n\;.
\ee
Of course, for the component fields in the $\theta$ and
$u, v$ expansion
of $\Psi^{(p,q)}$ one obtains ordinary reality conditions.

This is an appropriate place to comment on the relation to the
projective
$(4,4)\;\;2D$ superspace formalism \cite{{GHR},{BLR},{RSS}}.

Bearing
some formal resemblances to the latter, the harmonic
superspace approach differs in a number of important aspects.
This mainly regards the
treatment of extra bosonic variables which are present in both
approaches. In the projective $2D$ superspace formalism they
form two sets of complex
variables with respect to which one takes contour integrals,
while in the harmonic superspace formalism they are $SU_L(2)/U_L(1)$
and $SU_R(2)/U_R(1))$ spinor harmonics
$u^{(\pm 1)}_i\;,\;v^{(\pm 1)}_a$, and all the involved fields
are assumed to be decomposable into the harmonic series in these
variables.
The harmonic
variables $u, \;v$ represent the left and right spheres
$SU(2)_{L,R}/U_{L,R}(1)$ in a parametrization-independent
way, while the complex coordinates of the projective superspace
can be
viewed as particular parametrizations of the same spheres. In the
projective superspace approach, when doing contour integration,
there arises an uneasy problem of how to choose the relevant
integration contours. There is no such problem in the harmonic
superspace
approach where the integral over additional variables is
understood as
the double harmonic integral on the product
$SU_L(2)/U_L(1)\otimes SU(2)_R/U_R(1)$. It can be
defined by the rules \cite{GIKOS}
\be
\int du \;1 = 1\;,\;\; \int du \;u^{(1)}_{(i_1}...u^{(1)}_{i_k}
u^{(-1)}_{i_{k+1}}... u^{(-1)}_{i_n)} = 0
\label{int}
\ee
(for $v$ integration the rules are the same). The basic feature of the
harmonic superspace approach is
that the harmonic variables are treated on equal footing with
other superspace coordinates: with respect to them one not only
integrates, but also differentiates, they essentially enter into
the formulas relating central and analytic bases, they are
responsible for
the presence of an infinite number of auxiliary fields in
unconstrained analytic $(4,4)$ superfields, etc.

\vspace{0.4cm}
\noindent{\bf 2.4 Two N=4 SU(2) superconformal groups.} As the last
topic of this Section we will discuss realizations
of two $N=4\;\; SU(2)$ superconformal
groups in the analytic $(4,4)$ superspace (\ref{anharm}). To know
them will be very important for our further purposes.

Both these
$N=4\;\; SU(2)$
superconformal groups consist of
two commuting
branches independently acting in the left and right analytic
subspaces, so it will be sufficient to consider their action, say,
in ${\bf AL}^{(1+2|2)}$.

The existence of two different $N=4\;\;SU(2)$ superconformal
groups in ${\bf AL}^{(1+2|2)}$ is related to the fact that
the most general superconformal group which can be defined in
this superspace is the so called ``large'' $N=4$ superconformal
group with $SO(4)\times U(1) \sim SU(2)\times SU(2)\times U(1)$
affine Kac-Moody subgroup in its bosonic sector
\cite{{Ademollo},{Schoutens},{Belg2},{IKL2}}. The $N=4\;\;SU(2)$
superconformal groups in question are two different subgroups of
this ``large'' superconformal group, each including one of two
$SU(2)$ factors of the $SO(4)$ just mentioned. Transformations
of the ``large'' supergroup on the coordinates of
${\bf AL}^{(1+2|2)}$ have been already given in \cite{DelSok}.
We present the coordinate realizations of its two $N=4\;\; SU(2)$
subgroups separately, as they have essentially different
implications in the sigma models we
are going to discuss. Our way of deducing these realizations and
parametrizing them slightly differ from the one adopted in
\cite{DelSok}.

First of these superconformal groups acts on all coordinates of
${\bf AL}^{(1+2|2)}$ including
the harmonic ones (for brevity we omit the light-cone indices):
\bea
\delta_I\; z &=& \Lambda_I(\zeta_L, u)\;,\;
\delta_I \;\theta^{(1,0)} \;=\; \Lambda^{(1,0)}_I (\zeta_L, u)
\;,\;
\delta_I \;\bar \theta^{(1,0)} \;=\; \bar \Lambda^{(1,0)}_I
(\zeta_L, u)\;, \nonumber \\
\delta_I\; u^{(1)}_i &=& \Lambda^{(2,0)}_I u^{(-1)}_i\;,\;
\delta_I\; u^{(-1)}_i \;=\; 0
\label{scgI}
\eea
and is fully determined by the requirement that the harmonic
derivative $D^{(2,0)}$ transforms as
\be
\delta_I\; D^{(2,0)} = - \Lambda^{(2,0)}_I D^{(0,0)}_u\;.
\label{baseqI}
\ee
{}From this condition one obtains the relations \cite{DelSok}
\bea
D^{(2,0)} \Lambda_I &=& 2i\;(\bar \theta^{(1,0)}
\Lambda^{(1,0)}_I -
\theta^{(1,0)} \bar \Lambda^{(1,0)}_I)\;, \nonumber \\
D^{(2,0)} \Lambda^{(1,0)}_I &=& \Lambda^{(2,0)}_I
\theta^{(1,0)}\;, \;\;
D^{(2,0)} \bar \Lambda^{(1,0)}_I \;=\; \Lambda^{(2,0)}_I
\bar \theta^{(1,0)}\;,
D^{(2,0)} \Lambda^{(2,0)}_I = 0
\eea
which have a simple general solution via the constrained analytic
function
$a (\zeta_L, u)$
\bea
\Lambda_I &=& a - {1\over 2} \partial^{(-2,0)} D^{(2,0)} a \;, \;
\Lambda^{(1,0)}_I \;=\; {i \over 4} \frac{\partial}{\partial \bar
\theta^{(1,0)}}
D^{(2,0)} a \;,\;\nonumber \\
\bar \Lambda^{(1,0)}_I &=& -{i \over 4} \frac{\partial}{\partial
\theta^{(1,0)}}
D^{(2,0)} a \;,\;
\Lambda^{(2,0)}_I = - {1\over 2} D^{(2,0)} \partial_{z}a \;\equiv \;
D^{(2,0)} \Lambda^{(0,0)}_I\;,
\label{scgIdet}
\eea
\be
(D^{(2,0)})^2 a (\zeta_L, u) = 0 \;.
\label{aconstr}
\ee
The explicit form of $a (\zeta_L, u)$ is as follows
\bea
a(\zeta_L, u) &=& a_0(z) + a_0^{(ij)}(z) u^{(1)}_i u^{(-1)}_j +
\theta^{(1,0)} \xi^i (z)
u^{(-1)}_i + \bar \theta^{(1,0)} \bar \xi^{i} (z) u^{(-1)}_i
\nonumber \\
&&- 2
i \theta^{(1,0)} \bar \theta^{(1,0)} \partial_z a_0^{(ij)}(z)
u^{(-1)}_i u^{(-1)}_j \;.
\label{a}
\eea
Here $a_0 (z), \xi^i (z), \bar \xi^i (z), \partial_z a_0^{(ij)} (z)$
are, respectively, parameters of conformal, supersymmetry
and $SU(2)$ Kac-Moody
transformations forming the $N=4\;\; SU(2)$ superconformal group.
Notice
that the Kac-Moody transformation parameter enters $a$ via its
``prepotential'' $a^{(ij)}$. It is an easy exercise to check that the
transformations (\ref{scgI}, \ref{scgIdet}) preserve the
${\bf AL}^{(1+2|2)}$ integration measure
\bea
\mu^{(-2,0)} &\equiv & [d\zeta]^{(-2,0)}du \;=\;
dz\; d \theta^{(1,0)}\;
d\bar
\theta^{(1,0)}\; du\;, \nonumber \\
\delta_I \;\mu^{(-2,0)} &=& \left( \partial_z \Lambda_I +
\partial^{(-2,0)}
\Lambda^{(2,0)}_I - \frac{\partial \Lambda^{(1,0)}_I}{\partial
\theta^{(1,0)}}
- \frac{\partial \bar \Lambda^{(1,0)}_I}{\partial \bar \theta^{(1,0)}}
\right)\;
\mu^{(-2,0)} = 0\;.
\label{measI}
\eea
Note that the above analytic superspace realization of
$N=4\;\;SU(2)$ superconformal group
is basically of the same form as that of $N=2\;\;4D$ superconformal
group in the corresponding harmonic
analytic superspace \cite{essay}.

The second $N=4\;\; SU(2)$ superconformal group has no direct
analog in $N=2\;\;4D$ case. It does not affect harmonic
variables
\bea
\delta_{II}\; z &=& \Lambda_{II}(\zeta_L, u)\;,\;
\delta_{II}\; \theta^{(1,0)} \;=\; \Lambda^{(1,0)}_{II}
(\zeta_L, u)\;,\;
\delta_{II}\; \bar \theta^{(1,0)} \;=\;
\bar \Lambda^{(1,0)}_{II} (\zeta_L, u)\;, \nonumber \\
\delta_{II}\; u^{(\pm 1)}_i &=& 0
\label{scgII}
\eea
and is fully determined by requiring $D^{(2,0)}$ to be invariant
\be
\delta_{II}\; D^{(2,0)} = 0\;.
\label{baseqII}
\ee
The latter equation implies
\bea
D^{(2,0)} \Lambda^{(1,0)}_{II} &=& D^{(2,0)} \bar
\Lambda^{(1,0)}_{II}
\;=\; 0\;, \nonumber \\
D^{(2,0)} \Lambda_{II} &=& 2i\;(\bar \theta^{(1,0)} \Lambda^
{(1,0)}_{II} -
\theta^{(1,0)} \bar \Lambda^{(1,0)}_{II})\;.
\label{relII}
\eea
Combining $\theta^{(1,0)}, \bar \theta^{(1,0)}$ into a doublet of an
extra $SU(2)$ (recall (\ref{quartet})),
\be
\{ \theta^{(1,0)}, \bar \theta^{(1,0)} \} \equiv \{ \theta^{(1,0)\;
\underline{i}} \}\;,
\label{quartet1}
\ee
one can write the general solution to (\ref{relII}) as
\bea
\Lambda^{(1,0)\;\underline{i}}_{II} (\zeta_L, u) &=&
\lambda^{k \underline{i}}(z) u^{(1)}_k +
\theta^{(1,0)\;\underline{l}}\;
(\lambda^{\;\;\;\underline{i})}_{(\underline{l}}
-\delta^{\underline{i}}_{\underline{l}} \partial_z \lambda (z))
-2i\;\theta^{(1,0)\;\underline{t}}\;\theta^{(1,0)}_{\underline{t}}
\partial_z \lambda^{k \underline{i}}(z) u^{(-1)}_k\;,\nonumber \\
\Lambda_{II} (\zeta_L, u) &=& \lambda (z) + 2i\; \theta^{(1,0)}_
{\underline{k}}
\lambda^{i\underline{k}}(z) u^{(-1)}_i\;.
\eea
Here $\lambda(z), \lambda^{(\underline{i} \underline{k})}(z),
\lambda^{k \underline{i}}(z)$ are, respectively, the parameters of
conformal, second $SU(2)$ Kac-Moody and supersymmetry
transformations. The analytic superspace integration measure
is also preserved by this $N=4\;\; SU(2)$ superconformal group
\be
\delta_{II}\; \mu^{(-2,0)} = 0\;.
\label{measII}
\ee

Finally, we wish to mention once more that these two
superconformal groups do
not commute; their closure is the ``large'' $N=4$ superconformal
group. We will not discuss here the detailed structure of
this closure.

\setcounter{equation}{0}
\section{(4,4) sigma models in harmonic superspace}
The main reason why we applied to the $(4,4)$ harmonic superspace
formalism was the hope to construct, within its framework, an
off-shell
formulation of general $(4,4)$ supersymmetric sigma models with
torsion and, as a special subclass of the latter, $(4,4)$
superextensions of the group manifold WZNW sigma models.
As a first step in
approaching our ultimate aim, in this Section we
rewrite in the harmonic superspace the general $(4,4)$ sigma models
with two mutually commuting sets of complex structures.

\vspace{0.4cm}
\noindent{\bf 3.1 Harmonic superspace description of (4,4)
twisted multiplet.}
In $(2,2)$ superspace the
$(4,4)$ sigma models with mutually commuting left and right complex
structures are described by an action of
paired chiral and twisted
chiral superfields \cite{{GHR},{RSS},{RASS}}. These pairs comprise
the $(4,4)$ twisted
\cite{GHR} (or analytic \cite{IK2}) multiplet, so the $(4,4)$
superspace
form of the sigma model action in question should coincide
with a general action of $(4,4)$ superfields representing
the $(4,4)$ twisted multiplets. It turns out that this action
admits a natural formulation in the analytic harmonic superspace
${\bf AS}^{(1+2,1+2|2,2)}$ (\ref{anharm}).

We start by recalling how this multiplet is described in
conventional $(4,4)$ superspace.

It is represented by a real
quartet superfield $q^{ia}(Z_L, Z_R), (q^{ia})^\dagger =
\epsilon_{ik}\epsilon_{ab} q^{kb}$, subject to the following
irreducibility conditions \cite{{IK2},{GHR},{IKL},{Sieg}}
\footnote{The constraints in this form have been
given for the first time in \cite{IK2}.}
\be
D_+^{(j} \;q^{i)a} = \overline{D}_+^{(j} \;q^{i)a} = 0\;, \;\;
D_-^{(b} \;q^{ia)} = \overline{D}_-^{(b} \;q^{ia)} = 0\;.
\label{twistcon}
\ee
These constraints leave in $q^{ia}$  $8+8$ independent field
components
that is just the off-shell field content of the $(4,4)$
twisted multiplet
(it reduces to $4+4$ on shell).

Let us now convert the $SU(2)$ indices of $q^{ia}$ and spinor
derivatives
in (\ref{twistcon}) with the harmonics $u^{(1)}_i\;,\;v^{(1)}_a$
in order to
rewrite (\ref{twistcon}) in the following equivalent form
\bea
D^{(1,0)} q^{(1,1)}&=& \overline{D}^{(1,0)}
q^{(1,1)} \;=\; 0\;, \nonumber \\
D^{(0,1)} q^{(1,1)} &=& \overline{D}^{(0,1)} q^{(1,1)}
\;=\; 0\;,
\label{harmcon}
\eea
where the involved projections of the spinor derivatives are
defined in
eq. (\ref{covdan}) and
\be
q^{(1,1)} (Z_L, Z_R, u, v) \equiv q^{ia}(Z_L, Z_R) u_i^{(1)}
v_a^{(1)}\;,
\;\;\;\widetilde{q^{(1,1)}} = q^{(1,1)}\;.
\label{qu11}
\ee
The homogeneity property (\ref{qu11}) can be equivalently
reexpressed as the harmonic
constraints \cite{GIKOS}
\be
D^{(2,0)} q^{(1,1)} = D^{(0,2)} q^{(1,1)} = 0\;,
\label{harmcon2}
\ee
after which, comparing (\ref{harmcon}) with eqs. (\ref{ancon}), one
concludes that the $(4,4)$ twisted multiplet is represented in
the harmonic superspace by a real analytic superfield
$$
q^{(1,1)}= q^{(1,1)} (\zeta_L, \zeta_R, u,v)
$$
which obeys the harmonic constraints (\ref{harmcon2}).

Using the analytic superspace form of $D^{(2,0)}, D^{(0,2)}$,
eq. (\ref{anhder}), one can solve (\ref{harmcon2}) and find
the component structure of $q^{(1,1)}$
\bea
q^{(1,1)}(\zeta_L,\zeta_R,u,v) &=& q^{ia}(z)u^{(1)}_i v^{(1)}_a +
2 \theta^{(1,0)} \psi_+^a (z) v^{(1)}_a \nonumber \\
&&+
2 \bar \theta^{(1,0)} \bar \psi_{+\; a}(z)
v^{(1)a} +
2 \theta^{(0,1)} \chi_-^i (z) u^{(1)}_i \nonumber \\
&& + 2 \bar \theta^{(0,1)} \bar \chi_{-\; i} (z) u^{(1)i} - 2i\;
\theta^{(1,0)}\bar \theta^{(1,0)} \partial_{++} q^{ia}(z) u^{(-1)}_i
v^{(1)}_a \nonumber \\
&& - 2i\;
\theta^{(0,1)}\bar \theta^{(0,1)} \partial_{--} q^{ia}(z) u^{(1)}_i
v^{(-1)}_a + 2 \theta^{(1,0)}\theta^{(0,1)} F(z) \nonumber \\
&& - 2 \bar \theta^{(1,0)}\bar \theta^{(0,1)} \bar F(z)
+ 2 \theta^{(1,0)} \bar \theta^{(0,1)} L(z) +
2 \bar \theta^{(1,0)} \theta^{(0,1)} \bar L(z) \nonumber \\
&& -4i\; \theta^{(1,0)} \theta^{(0,1)}\bar \theta^{(0,1)}
\partial_{--} \psi_+^a (z) v^{(-1)}_a \nonumber \\
&&+ 4i\; \bar \theta^{(1,0)} \bar \theta^{(0,1)} \theta^{(0,1)}
\partial_{--} \bar \psi_{+\; a} (z) v^{(-1)a} \nonumber \\
&& -4i\; \theta^{(0,1)} \theta^{(1,0)} \bar \theta^{(1,0)}
\partial_{++}\chi_-^i(z) u^{(-1)}_i \nonumber \\
&&+ 4i\; \bar \theta^{(0,1)} \bar \theta^{(1,0)} \theta^{(1,0)}
\partial_{++} \bar \chi_{-\; i} (z) u^{(-1)i} \nonumber \\
&& - 4\; \theta^{(1,0)}\bar \theta^{(1,0)} \theta^{(0,1)}
\bar \theta^{(0,1)}
\partial_{++}\partial_{--} q^{ia}(z) u^{(-1)}_i v^{(-1)}_a\;,
\label{compon}
\eea
where ``bar'' on the fields means ordinary complex conjugation
and some
numerical factors have been inserted for further convenience. We
see that the fields $q^{ia}, \psi_+^a, \bar \psi_{+}^a,
\chi_-^i, \bar \chi_{-\;i}$ and $F, \bar F, L, \bar L $ have
appropriate dimensions
to represent, respectively, physical and auxiliary
degrees of freedom. Note a formal similarity of the $(4,4)$ harmonic
constraints
(\ref{harmcon2}) to the constraints defining the $N=2$ tensor multiplet
in the harmonic $N=2\;\;4D$ superspace \cite{GIO1}. The crucial
difference between the two types of constraints is
that the latter implies a differential condition for a vector
component of the relevant superfield, requiring it to be
divergenceless,
while this is not the case for the constraints (\ref{harmcon2}).

As a last topic of this Subsection we discuss the transformation
properties
of $q^{(1,1)}$ under two $N=4\;\; SU(2)$ superconformal
groups defined
in Subsect.2.4. These transformation laws are uniquely fixed by the
requirement of preserving the harmonic constraints
(\ref{harmcon2}) and
turn out to be very simple (we again omit the light-cone indices)
\bea
\delta_I \;q^{(1,1)} (\zeta_L, \zeta_R, u, v) &\simeq &
q^{(1,1)'} (\zeta_L', \zeta_R, u', v) -
q^{(1,1)}(\zeta_L, \zeta_R, u, v)
\nonumber \\
&=& \Lambda^{(0,0)}_I q^{(1,1)} (\zeta_L, \zeta_R, u, v)
\nonumber \\
&=& -{1\over 2}\left( \partial_{z} a (\zeta_L, u) \right)
q^{(1,1)} (\zeta_L, \zeta_R, u, v)
\;,
\label{supquI} \\
\delta_{II} \;q^{(1,1)} (\zeta_L, \zeta_R, u, v) &\simeq &
q^{(1,1)'}(\zeta_L', \zeta_R, u, v) - q^{(1,1)}
(\zeta_L, \zeta_R, u, v)
\;=\;0\;.
\label{supquII}
\eea
The transformation rules with respect to the right light-cone
branches
of these superconformal groups are given by similar formulas.

The
basic difference between the realizations I and II lies in the action
of the $SU(2)$ affine Kac-Moody subgroup: in the case I it acts
both on the physical bosonic
and fermionic fields as rotations of their indices $i$, $a$ while
in the case II
it does not affect the physical bosons $q^{ia}$ at all and acts
only on fermions, mixing $\psi$ with $\bar \psi$ and $\chi$
with $\bar \chi$. The
auxiliary fields are scalars with respect to the $SU(2)$ Kac-Moody
subgroup of the realization I and split into a singlet and triplet
with respect to an analogous subgroup of the realization II. All these
properties become manifest in the ``quartet'' notation
(\ref{quartet1}). For instance, four auxiliary fields terms in
(\ref{compon}) are combined into the single term
\bea
2\theta^{(1,0)\;\underline{i}} \theta^{(0,1)\;\underline{a}}\;
F_{\underline{i}\;\underline{a}}\;,
\; \; \;\;\;\;F_{\underline{i}\;\underline{a}} &=&
\left( \begin{array}{cccc}
F & L \\
\bar L & -\bar F
\end{array} \right)\;.
\label{quartF}
\eea

As a useful example of the component transformation properties
we explicitly give the transformation rule of the bosonic field
$q^{ia}$ under the $SU(2)_I$ Kac-Moody subgroup
\be
\delta_{SU(2)_I} \;q^{i a}(z^{++}, z^{--}) =
{1\over 2} \left( \partial a_0^{(ik)} (z^{++}) \right)
q_{k}^{\;a}(z^{++}, z^{--})\;.
\label{KMI}
\ee

Note that the
superconformal transformations of the constrained superfield
$q^{ia}(Z_L,Z_R)$ representing the $(4,4)$ twisted multiplet in the
conventional
$(4,4)$ superspace were given in \cite{GI}. These look much more
complicated compared to the $(4,4)$ analytic superfield
ones (\ref{supquI}), (\ref{supquII}).

\vspace{0.4cm}
\noindent{\bf 3.2 General action of the superfields $q^{(1,1)}$.}
By the
dimensionality reasoning and keeping in mind the requirement of
conservation of the $U(1)$
charges, it is straightforward to write the most general action of
self-interacting  superfields $q^{(1,1)\;M}$ ($M =1,2, ...$)
\be
S_q = \int \mu^{(-2,-2)}\; {\cal L}^{(2,2)} (q^{(1,1)\;M}
(\zeta_L,\zeta_R,u,v),\;
u,\; v)\;.
\label{genaction}
\ee
Here
$$
\mu^{(-2,-2)} \equiv \mu^{(-2,0)} \mu^{(0,-2)} =
d^2z\; d^2\theta^{(1,0)}\;d^2 \theta^{(0,1)}\;du\;dv
$$
is the measure of integration over the analytic $(4,4)$ superspace.
The dimensionless analytic superfield Lagrangian
${\cal L}^{(2, 2)} (q^{(1,1)\;M}, u^{(\pm 1)}_i, v^{(\pm 1)}_a) $
bears in general an arbitrary dependence on its arguments, the only
restriction being a compatibility
with the external $U(1)$ charges $(2,\; 2)$ of the Lagrangian. The
free action of $q^{(1,1)\;M}$ is given by
\be
S_{q}^{free} \sim \int \mu^{(-2,-2)}\; q^{(1,1)\;M}\;
q^{(1,1)\;M}\;,
\label{free}
\ee
so for consistency we are led to assume
\be
\mbox{det} \left( \frac{\partial^2 {\cal L}^{(2,2)}}{\partial
q^{(1,1)\;M}
\partial q^{(1,1)\;N}} \right)\;|_{q^{(1,1)} = 0} \neq  0\;.
\ee
For completeness, we also add the constraints
on $q^{(1,1)\;M} (\zeta_L,
\zeta_R,u,v)$
\be
D^{(2,0)}q^{(1,1)\;M} = D^{(0,2)}q^{(1,1)\;M} =0\;.
\label{constrM}
\ee

It is straightforward to substitute the component expansion of
$q^{(1,1)}$, (\ref{compon}), into (\ref{genaction}), to
integrate
over $\theta$'s and to obtain the component form of the action.
It is instructive to give here its physical and auxiliary
bosons parts,
with all fermions omitted. These pieces can be written as follows
\be
S_{phb} = 2\int d^2 z\; \{ G_{Mia\;Njb} (q) \;\partial_{++}
q^{ia\;M}
\partial_{--}q^{jb\;N} + B_{Mia\;Njb}(q)\; \partial_{++}
q^{ia\;M}
\partial_{--}q^{jb\;N} \}\;,
\label{compbosph}
\ee
\be \label{compbosaux}
S_{auxb} = 4 \int d^2 z\; G_{M\;N} (q)\; (F^M \bar{F}^N +
L^M \bar{L}^N)\;,
\ee
where
\bea
G_{Mia\;Njb} (q) &=& \int du dv\; g_{M\;N}(q^{(1,1)}_0, u,
v)\; \epsilon_{ij} \epsilon_{ab}\;, \label{metr} \\
B_{Mia\;Njb}(q) &=& \int du dv\; g_{M\;N}(q^{(1,1)}_0, u,
v)[\epsilon_{ij} v^{(1)}_{(a} v^{(-1)}_{b)} -
\epsilon_{ab} u^{(1)}_{(i} u^{(-1)}_{j)}]\;, \label{tors} \\
G_{M\;N} (q) &=& \int du dv\; g_{M\;N}(q^{(1,1)}_0, u, v)\;.
\label{auxfun} \\
g_{M\;N}(q^{(1,1)}_0, u, v) &=&
\frac{\partial^2 {\cal L}^{(2,2)}}{\partial q^{(1,1)\;M}
\partial q^{(1,1)\;N}}|_{\theta = 0}\;,
\label{metr2}
\eea
where $q_0^{(1,1)} \equiv q^{(1,1)} |_{\theta = 0}$. The
objects $G_{Mia\;Njb} (q)$, $B_{Mia\;Njb}(q)$ are,
respectively, symmetric and antisymmetric under the
simultaneous permutation of the
indices $M \leftrightarrow N, \; i \leftrightarrow j, \;
a\leftrightarrow b$
and so they can be identified with the metric and torsion
potential on the target space.

Sometimes it is advantageous to represent the second
term in (\ref{compbosph}) through the torsion field strength.
It is introduced by the standard expression
\be
H_{Mia\;Njb\;Tkd} = \frac{\partial B_{Njb\;Tkd}}
{\partial q^{ia\;M}} +
\frac{\partial B_{Mia\;Njb}}{\partial q^{kd\;T}} +
\frac{\partial B_{Tkd\;Mia}}{\partial q^{jb\;N}}\;,
\label{hgen}
\ee
and is totally antisymmetric with respect to permutations of
the triples $Mia, Njb, Tkd$.
Letting $q^{ia\;M}$ depend on an extra parameter
$t$, with $q^{ia\;M}(t,z)|_{t=1} \equiv q^{ia\;M}(z),\;
q^{ia\;M}(t,z)|_{t=0} = \epsilon^{ia}$, one can locally
rewrite the torsion term as
\be
B_{Mia\;Njb}\; \partial_{++} q^{ia\;M} \partial_{--} q^{jb\;N} =
\int^1_0 dt\; H_{Mia\;Njb\;Tkd}\; \partial_t q^{ia\;M}
\partial_{++} q^{jb\;N} \partial_{--} q^{kd\;T}\;.
\label{torsHg}
\ee
For $B_{Mia\;Njb}$ given by eq. (\ref{tors}),
$H_{Mia\;Njb\;Tkd}$ is reduced to
\be
H_{Mia\;Njb\;Tkd}(q) = \partial_{(Mid} \; G_{N\;T)}(q) \;
\epsilon_{ab}
\epsilon_{jk} + \partial_{(Mka}\; G_{N\;T)}(q)\;
\epsilon_{db} \epsilon_{ij}\;,
\label{exprHg}
\ee
where $\partial_{Mid} \equiv \partial / \partial q^{id\;M}$
and symmetrization is with respect to indices $M,N,T$.
Note that all the fermionic terms in the action
(\ref{genaction}) are also
expressed through the function $G_{M\;N}(q)$ and its
derivatives.

Thus we see that in the harmonic superspace formalism all
the target geometry objects associated with the off-shell
sigma model action (\ref{genaction})
are expressed in terms of the metric $G_{M\;N}(q)$ which is given
by a double harmonic integral of the
second derivative of the single function, the analytic superspace
Lagrangian
${\cal L}^{(2,2)}(q^{(1,1)}, u, v)$. A similar representation
for these geometric objects has been
obtained earlier in the projective superspace approach \cite{GHR},
with contour integrals instead of the harmonic ones.
The quantity
${\cal L}^{(2,2)}$ is an analog of the hyper-K\"ahler
potential \cite{GIOSgeo}. It would be tempting to find out the
appropriate geometric setting within which the representation
(\ref{metr}) - (\ref{metr2}),
(\ref{exprHg}) would follow from certain first principles,
like this has been done, e.g., for the hyper-K\"ahler geometry in
\cite{GIOSgeo}, and for the geometries of sigma models
with heterotic supersymmetry in \cite{DKS}. There,
by solving the defining
constraints on the curvature and torsion, expressions for
all geometric quantities through a few unconstrained potentials
have been obtained.

As the last remark we note that the action
(\ref{genaction}) with {\it arbitrary} ${\cal L}^{(2,2)}$ respects
invariance under the second of two  $N=4 \;\; SU(2)$ superconformal
groups realized in the analytic superspace
(eqs. (\ref{scgII}) - (\ref{measII}), (\ref{supquII})).
As for
the superconformal group defined by eqs.
(\ref{scgI}) - (\ref{measI}),
(\ref{supquI}), in general it does not constitute an
invariance group of the
action. Even in the free case (\ref{free}) this symmetry
is broken. An action possessing this superconformal symmetry
will be presented in the next Section.

\setcounter{equation}{0}
\section{$SU(2)\times U(1)$ WZNW sigma model in harmonic
superspace}

In this Section we show that the requirement of invariance
under the
$N=4\;\; SU(2)$ superconformal group (\ref{scgI}) - (\ref{measI}),
(\ref{supquI}) uniquely fixes the $q^{(1,1)}$ sigma model action
to be that of
$N=4 \;\;SU(2)\times U(1)$ WZNW sigma model \cite{{IKL},{RSS},{GI}}.

\vspace{0.4cm}
\noindent{\bf 4.1 $N=4\; \;SU(2)$ superconformally invariant
$q^{(1,1)}$ action.} Let us specialize to a single
superfield $q^{(1,1)}$ and
construct for it an action invariant under the superconformal
group defined by eqs. (\ref{scgI}) - (\ref{measI}),
(\ref{supquI}).

As was already mentioned, the free action (\ref{free}) does not
respect this superconformal invariance (even its rigid scale and
$SU(2)$ subsymmetries), so the invariant action should
necessarily include self-interaction terms. To find its precise form,
we apply the procedure which has been employed earlier in \cite{GIO1}
for constructing the
action of improved $N=2\;\; 4D$ tensor multiplet in the analytic
harmonic $N=2\;\; 4D$ superspace. Namely, we split $q^{(1,1)}$ as
\bea
q^{(1,1)} &=& \hat{q}^{(1,1)} + c^{(1,1)}\;,\;\; c^{(1,1)} \equiv
c^{ia}u^{(1)}_i v^{(1)}_a\;,  \label{split} \\
D^{(2,0)}\hat{q}^{(1,1)} &=& D^{(0,2)}\hat{q}^{(1,1)} \;=\; 0\;,
\label{constrhat}
\eea
with $c^{ia}$ being a quartet of arbitrary constants, and represent
the sought analytic superspace action as a series in
$\hat{q}^{(1,1)}$
\be
S_{sc} = \int \mu^{(-2,-2)}\; \sum^{\infty}_{n=2} b_n
(\hat{q}^{(1,1)})^{n} (c^{(-1,-1)})^{n-2}\;.
\label{prob}
\ee
Here the appropriate degrees of
$c^{(-1,-1)} = c^{ia}u^{(-1)}_i v^{(-1)}_a$ have been
inserted for the balance of $U(1)$ charges. The newly
introduced analytic
superfield $\hat{q}^{(1,1)}$ transforms inhomogeneously under the
superconformal transformations (\ref{scgI}) - (\ref{measI}),
(\ref{supquI})
(it will be sufficient to consider only the left light-cone
branch of
the whole superconformal group)
\be
\delta_I\; \hat{q}^{(1,1)} = \Lambda^{(0,0)}_I \hat{q}^{(1,1)} +
\Lambda^{(0,0)}_I c^{(1,1)} - \Lambda^{(2,0)}_I c^{(-1,1)}\;,\;
c^{(-1,1)} = c^{ia}u^{(-1)}_i v^{(1)}_a\;,
\label{transfhat}
\ee
so there arises an opportunity to achieve the invariance of
(\ref{prob})
by requiring that the variations of the terms of different order in
$\hat{q}^{(1,1)}$ cancel each other up to full harmonic derivatives.
Namely, we take into account the invariance of the integration
measure and then demand the homogeneous part of the variation of
the second order term to be cancelled by the inhomogeneous part
of the variation of
the third order term, etc (the inhomogeneous part of the
variation of the second order term
is a full harmonic derivative in its own right upon using
the constraints (\ref{constrhat})). In the process, one exploits
the
defining constraints (\ref{constrhat}) and the identities
\bea
c^{(1,-1)} c^{(-1,1)} &=&  c^{(-1,-1)} c^{(1,1)}  -
{1\over 2} c^{2} \;, \;\;
c^2 \equiv c^{ia}c_{ia}\;, \nonumber \\
D^{(2,0)} D^{(0,2)} \;(c^{(-1,-1)})^n &=& n^2 (c^{(-1,-1)})^{n-1}
\;c^{(1,1)}
-{1\over 2} c^2 n(n-1) (c^{(-1,-1)})^{n-2}\;,
\eea
which follow from the completeness relations
$$
u^{(1)}_i u^{(-1)}_j - u^{(-1)}_i u^{(1)}_j = \epsilon_{ij}\;, \;
v^{(1)}_a v^{(-1)}_b - v^{(-1)}_a v^{(1)}_b = \epsilon_{ab}
$$
and the definition of $D^{(2,0)},\;D^{(0,2)}$. Doing in this way, one
finally proves that the action (\ref{prob}) is invariant provided the
following
recurrence relations between the coefficients $b_n$ hold
\be
b_{n+1} = - {2\over c^2}\; \left( \frac{n^2}{n^2 - 1} \right)\; b_n\;,
\label{rec}
\ee
whence one finds
\be
b_n = 2 \left( -{2\over c^2} \right)^{n-2} \frac{n-1}{n}\; b_2\;.
\label{recI}
\ee
Now, introducing
\be
X \equiv 2\left( \frac{c^{(-1,-1)}\hat{q}^{(1,1)}}{c^2} \right)\;,
\ee
it is straightforward to show that the series in (\ref{prob})
is summed up to the expression
\bea
S_{sc} &\equiv &
- 2b_2 \int \mu^{(-2,-2)} {\cal L}^{(2,2)}_{sc}
(\hat{q}^{(1,1)}, u,v)
=  -2b_2 \int \mu^{(-2,-2)} \;\hat{q}^{(1,1)} \hat{q}^{(1,1)}
\left( \frac{\mbox{ln}(1+X)}{X} \right)'
\nonumber \\
&=& -2b_2 \int \mu^{(-2,-2)} \;\hat{q}^{(1,1)} \hat{q}^{(1,1)}
\left( \frac{1}{(1+X)X} -\frac{\mbox{ln}(1+X)}{X^2} \right)\;.
\label{confact}
\eea
The action (\ref{confact}) is the sought superconformally invariant
$q^{(1,1)}$ action. We will prove later (in Subsect.4.2) that
it is just
the off-shell action of the $N=4\;\; SU(2)\times U(1)$ WZNW sigma
model (actually this could be figured out already from the fact that
the $SU(2)$
Kac-Moody subgroup of the superconformal group in question acts on the
physical bosonic fields $q^{ia}(z)$ in the way just specific for the
realizations of Kac-Moody symmetries in WZNW models,
see eq. (\ref{KMI})).
Now we dwell on some peculiarities of this action.

Let us first mention that its superconformal invariance can be
checked directly, without expanding it in a series in
$\hat{q}^{(1,1)}$. After a straightforward
computation with making use of the following simple formula for
the variation
\be
\delta\; S_{sc} = 2b_2 \int \mu^{(-2,-2)}\; \hat{q}^{(1,1)}\; \delta
\hat{q}^{(1,1)} \frac{1}{(1+X)^2}\;,
\ee
one finds
\bea
\delta_I{\cal L}^{(2,2)}_{sc} &=& D^{(2,0)} (c^{(-1,1)}
\hat{q}^{(1,1)} \frac{1}{(1+X)^2} \Lambda^{(0,0)}_I) - D^{(0,2)}
(c^{(1,-1)} \hat{q}^{(1,1)} \frac{2+X}{(1+X)^2}
\Lambda^{(0,0)}_I) \nonumber \\
&+& 2 D^{(0,2)}\hat{q}^{(1,1)} c^{(1,-1)} \frac{1}{(1+X)^3}
\Lambda^{(0,0)}_I
- D^{(2,0)} \hat{q}^{(1,1)} c^{(-1,1)} \frac{1-X}{(1+X)^3}
\Lambda^{(0,0)}_I
\label{var}
\eea
By virtue of the constraints (\ref{constrhat}) this variation
reduces to full harmonic derivatives, thus ensuring the
invariance of
the action $S_{sc}$ (\ref{confact}). In the same way one checks
the invariance of (\ref{confact}) under the right $2D$ light-cone
branch of the first
$N=4 \;\; SU(2)$ superconformal group. Of course,
being a particular case of the action (\ref{genaction}),
(\ref{confact})
is manifestly invariant under the second type $N=4\;\; SU(2)$
superconformal transformations (\ref{scgII}) - (\ref{measII}),
(\ref{supquII}).

As a next comment we point out that the
action (\ref{confact}) can be
uniquely restored (the relations (\ref{rec}),
(\ref{recI}) can
be deduced) merely by requiring it to be invariant under
some special
subgroups
of the left (or right) $N=4\;\; SU(2)$ superconformal group:
either under rigid
scale transformations with $\Lambda^{(0,0)}_I = \lambda\;,\;
\partial_z
\lambda = 0$, or under rigid $SU(2)_c$ transformations with
$\Lambda^{(0,0)}_I = \lambda^{(ij)} u^{(1)}_iu^{(-1)}_j\;,
\Lambda^{(2,0)} = \lambda^{(ij)} u^{(1)}_iu^{(1)}_j\;, \;
\partial_z \lambda^{(ij)} = 0$; considering the whole set of
$z$ dependent left and right superconformal transformations brings
nothing new in the proof of invariance.

Last two remarks concern the already mentioned analogy with the
improved $N=2\;\; 4D$ tensor multiplet.

The original Ansatz for the action
(\ref{prob}) and the transformation law (\ref{transfhat})
look very similar
to those used in ref. \cite{GIO1} (leaving aside the fact that the
$N=2\;\;4D$ superconformal group is finite-dimensional while
its $2D$
counterpart is infinite-dimensional). However, the relevant
recurrence relations between the coefficients $b_n$ and the
final expressions
for the superconformally invariant action radically differ.
The origin of this
difference lies in different superconformal properties of the
integration measure of the $N=2\;\;4D$ and $(4,4)\;\; 2D$ analytic
harmonic superspaces: in the former case it has the dimension
[$cm^2$] and is transformed by the relevant superconformal group,
while in the latter case it is dimensionless and superconformally
invariant. As a result, the superconformal invariance of the
action is achieved under different conditions on the
coefficients $b_n$.

A close analogy with the $N=2\;\; 4D$ tensor multiplet
action is retained in what concerns the constant $c^{ia}$.
In both
cases the presence of such a constant (an isotriplet one
in the $N=2\;\;4D$
case and an isoquartet one in the $(4,4)\;\; 2D$ case) is
inevitable in
the analytic superfield Lagrangian density, while the invariant
action, being rewritten
via the unshifted superfield, i.e. $q^{(1,1)}$
in the case at hand, does not depend on the specific choice of
this constant. This latter property is directly related to
the invariance of the action (\ref{confact}) under the rigid
scale and
$SU(2)_c$ subgroups of $N=4\;\;SU(2)$ superconformal group.

To demonstrate this, let us put
\be
b_2 = {1\over {4\sqrt{2}}}\; \frac{1}{c\;\kappa^2}\;, \;\;c
\equiv \sqrt{c^{ia}c_{ia}}\;,
\label{newconst}
\ee
and rewrite
the action (\ref{confact}) through $q^{(1,1)}$
\be
S_{sc} \equiv S_{sc} (q^{(1,1)}, c^{ia})\;.
\label{actqu}
\ee
Next, let us consider infinitesimal deformations of
(\ref{actqu}) under
some rigid dilatations and $SU(2)$ rotations of the
constant $c^{ia}$
\bea
\delta S_{sc} &\simeq & S_{sc}(q, c + \delta c) -
S_{sc}(q,c)\;, \nonumber \\
(a) \; \delta_1 c^{ia} &=& \alpha c^{ia}\;;\;\;\;
(b) \;\delta_2 c^{ia} \;=\;
\alpha^{(ik)} c_k^{\;a}\;.
\label{deform}
\eea
Keeping in mind eq.(\ref{newconst}), it is an easy exercise
to check that
these deformations are reduced, up to
full harmonic derivatives in the variation of the analytic
superfield Lagrangian ${\cal L}^{(2,2)}_{sc}$, to particular
$N=4\;\;SU(2)$ superconformal
transformations of $S_{sc}$ with the parameters
$$
(a) \;\Lambda^{(0,0)}_I = -\alpha\;; \;
(b) \; \Lambda^{(0,0)}_I = \alpha^{(ik)}u^{(1)}_i u^{(-1)}_k\;,\;
\Lambda^{(2,0)} = \alpha^{(ik)}u^{(1)}_i u^{(1)}_k\;.
$$
Hence, because of superconformal invariance of the action,
\be
(a)\;\delta_1 S_{sc} = 0\;; \;\;(b)\; \delta_2 S_{sc} = 0\;.
\ee

{}From first of these relations one concludes that the action
does not depend on the norm of the four-vector $c^{ik}$;
from now on, we choose
\be
c^{2} = 2 \Rightarrow b_2 = {1\over 8\kappa^2}\;,\; X =
c^{(-1,-1)}\hat
{q}^{(1,1)}\;,
\label{cnorm}
\ee-

{}From the second relation and an analogous relation which comes
from  considering an $SU(2)$ rotation of $c^{ia}$ in the
index $a$ it follows that the action does not depend
on the angular part of the four-vector $c^{ia}$ as well. So one
can put
\be
c^{ia} = \epsilon^{ia}\;.
\label{cpartic}
\ee

Though $c^{ia}$ drops out from the action, its presence in
${\cal L}^{(2,2)}_{sc}$ (even written through $q^{(1,1)}$) is
unavoidable. As was noticed in \cite{GIO1}, the presence of an
arbitrary isotriplet
constant in the analytic
superfield Lagrangian of the improved $N=2\;\; 4D$ tensor
multiplet has a deep topological meaning: this constant
parametrizes a Dirac-like
string of singularities appearing in the component action when the
latter is written
through the field strength of notoph \cite{deWit}. A meaning of the
constant $c^{ia}$ is somewhat more obscure. It seems to reflect an
ambiguity in resolving the torsion field strength 3-form
(which is closed but not exact in the case at hand) through the
2-form potential. Just
the latter enters into the component Lagrangian directly following
from ${\cal L}_{sc}^{(2,2)}$.
This interpretation is supported by the fact that the explicit
$c^{ik}$ dependence
in the torsion term of the Lagrangian disappears if one
writes this term through the torsion field strength which is
well defined globally, rather than through
the torsion  potential (see the next Subsection).

Keeping in mind eq. (\ref{cnorm}), the final form of the action
(\ref{confact}) is as follows
\bea
S_{sc} &=& -{1\over 4\kappa^2} \int \mu^{(-2,-2)}
{\cal L}^{(2,2)}_{sc}\;
\nonumber \\
{\cal L}^{(2,2)}_{sc} &=&
\hat{q}^{(1,1)} \hat{q}^{(1,1)} \left( \frac{1}{(1+X)X} -
\frac{\mbox{ln}(1+X)}{X^2} \right)
\;.\label{confact1}
\eea

\vspace{0.4cm}
\noindent{\bf 4.2 Passing to components.} In order to
demonstrate that the
action (\ref{confact1}) indeed describes the $N=4 \;\;SU(2)
\times U(1)$ WZNW model, we give here its component form and
show that it precisely coincides with the component
$N=4 \;\;SU(2)\times U(1)$ WZNW action \cite{{IKL},{GI}}.

Let us begin with the bosonic part of the action. It is given by a
sum of the physical and auxiliary bosonic fields terms defined
in eqs. (\ref{compbosph}) -
(\ref{metr2}). In the present case:
\bea
S_{sc}^{bos} &=& {1\over 2\kappa^2} \int d^2 z
\{ G(\hat{q})\; \partial_{++} q^{ia} \partial_{--} q_{ia} +
B_{ia\;jb}(\hat{q})\; \partial_{++} q^{ia} \partial_{--} q^{jb}
\nonumber \\
&& + 2 G(\hat{q})\;( F\bar F +  L\bar L ) \}
\label{compbossc} \\
G (\hat{q}) &=& -\int du dv\;
\frac{\partial^2 {\cal L}^{(2,2)}_{sc}}{\partial {q^{(1,1)}
\partial q^{(1,1)}}}|_{\theta = 0} \;=\;
\int du dv \; \frac{1-X}{(1+X)^3} \label{metrsc} \\
B_{ia\;jb} (\hat{q}) &=& \int du dv \; \frac{1-X}{(1+X)^3}
[\epsilon_{ij}
v^{(1)}_{(a}v^{(-1)}_{b)} -
\epsilon_{ab} u^{(1)}_{(i}u^{(-1)}_{j)}]\;.
\label{torssc}
\eea

It turns out that all the target geometry quantities present in the
Lagrangian (including its fermionic part) are eventually expressed
through
the single object $G(\hat{q})$ (\ref{metrsc}). It has been computed in
Appendix. It is a function of the original field
$q^{ia}(z)$ and it contains no explicit dependence on $c^{ia}$
\be
G(q) = 2 (q^{ia} q_{ia})^{-2} \equiv 2 \rho^{-2} \;.
\label{exprG1}
\ee
Parametrizing the $4\times 4$ physical bosons matrix $q^{ia}(z)$
as
\be
q^{ia} = e^{u(z)} \tilde{q}^{ia}(z)\;,
\label{newparam}
\ee
where $\tilde{q}^{ia}(z)$ is an unitary $SU(2)$ matrix,
\be
\tilde{q}^{ia} \tilde{q}^{j}_{\;a} = \epsilon^{ji}\;,
\tilde{q}^{ia} \tilde{q}_{i}^{\;b} = \epsilon^{ba}\;,
\ee
one finds that
\be
G(q) = e^{-2u}\;.
\label{exprG2}
\ee
So, the metric term in (\ref{compbossc}) is reduced to a sum
of the free Lagrangian of the field $u(z)$ and the Lagrangian
of the $SU(2)$ principal sigma model
$$
G(q) \;\partial_{++}q^{ia} \partial_{--}q_{ia} =
2 \partial_{++}u \partial_{--}u + \partial_{++}\tilde{q}^{ia}
\partial_{--}\tilde{q}_{ia}\;
$$

The last, torsion term in (\ref{compbossc}) needs a bit more
careful treatment. It is difficult to directly integrate
there over harmonic variables, the reason is that the method
of doing such integration
which we applied in \cite{GIO1} and in the Appendix requires
a manifest $SU(2)_c$ covariance of the relevant harmonic
integral, while the $SU(2)_c$ variation of the
integrand in the torsion term vanishes only modulo full $z$
derivatives.
To get round this difficulty, one can do as in Subsect.4.2
and rewrite the
torsion term via the field strength of the potential $B_{ia\;jb}$.
The totally antisymmetric (with
respect to permutations of pairs of the indices $ia,\;jb\;,...$)
torsion
field strength $H_{ia\;jb\;kd}$ defined by the general formula
(\ref{hgen}) in the given specific case is reduced to the simple
expression
\be
H_{ia\;jb\;kd} = \frac{\partial G(q)}{\partial q^{jd}}
\epsilon_{ab}
\epsilon_{ik} - \frac{\partial G(q)}{\partial q^{kb}}
\epsilon_{ad} \epsilon_{ij}\;,
\label{exprH}
\ee
or, with taking account of eq. (\ref{exprG1}),
\be
H_{ia\;jb\;kd} = 4 \rho^{-4} \left( q_{kb}\; \epsilon_{ad}
\epsilon_{ij}
- q_{jd}\; \epsilon_{ab} \epsilon_{ik} \right)\;.
\label{exprH1}
\ee
After substituting this expression into the torsion term
\be
B_{ia\;jb} \partial_{++} q^{ia} \partial_{--} q^{jb} =
\int^1_0 dt\; H_{ia\;jb\;kd}\; \partial_t q^{ia}
\partial_{++} q^{jb} \partial_{--} q^{kd}
\label{torsH}
\ee
and
passing to the parametrization (\ref{newparam}),
the r.h.s. of (\ref{torsH})
takes the form
\be
\int^1_0 dt\; \partial_t\tilde{q}_{ia}\;\tilde{q}_{jb}\; (
\partial_{++}\tilde{q}^{ib}\partial_{--}\tilde{q}^{ja} -
\partial_{++}\tilde{q}^{ja}\partial_{--}\tilde{q}^{ib})
\ee
which is the standard $SU(2)$ WZNW term.

Note that in this form WZNW term is well-defined globally,
it is also immediately
seen from the representation (\ref{exprH}), (\ref{exprH1})
that the field strength $H_{ia\;jb\;kd}$ transforms
as a tensor under the rigid $SU(2)_c$. As is explained in Appendix,
this is the main reason why this object depends only on unshifted
$q^{ia}$
and reveals no dependence on the constant $c^{ia}$. On the other
hand, the torsion potential $B_{ia\;jb}$ directly appearing in the
component action is defined only locally and it possesses no tensor
properties under $SU(2)_c$. By this reason it cannot be expressed
only in terms of $q^{(1,1)}$: one can check that the potential
inevitably includes a dependence on $c^{ia}$ which disappears
only after performing $z$ integration. So the presence of this
constant in $B_{ia\;jb}$ reflects the global uncertainty in resolving
$H_{ia\;jb\;kd}$ through $B_{ia\;jb}$\footnote{The off-shell action of
$N=4\;SU(2)\times U(1)$ WZNW
model in the conventional $(4,4)\;\;2D$ superspace from the beginning
contains an integral over extra $t$ \cite{GI}, so after passing to
components it yields just the $H$ form of the torsion term. On the
other hand, while the same model is formulated in terms of
$(2,2)\;\;2D$ superfields (chiral and twisted chiral ones) \cite{RSS},
the torsion term appears in its $B$ form, like in the $(4,4)$
harmonic superspace formulation.}.

Summing up the above contributions, one may write the final
expression for
the bosonic part of the action (\ref{confact1})
\bea
S_{sc}^{bos} &=& {1\over \kappa^2} \int d^2 z \;\{
\partial_{++}u \partial_{--}u + {1\over 2}
\partial_{++}\tilde{q}^{ia}
\partial_{--}\tilde{q}_{ia} \nonumber \\
&& +
{1\over 2} \int^1_0 dt \; \partial_t\tilde{q}_{ia}\;
\tilde{q}_{jb}\; (
\partial_{++}\tilde{q}^{ib}\partial_{--}\tilde{q}^{ja} -
\partial_{++}\tilde{q}^{ja}\partial_{--}\tilde{q}^{ib})
\nonumber \\
&& + e^{-2u} (F\bar F + L\bar L) \}\;.
\label{okonb}
\eea

Let us now apply to the fermionic sector. The fermionic part of the
component action
consists of three pieces
$$
S^{ferm}_{sc} = S_{4f} + S_{auxf} + S_{kinf}
$$
which correspond, respectively, to the term quartic in fermionic
fields, a term involving auxiliary
fields and the kinetic term of the component Lagrangian.
Explicitly, these are as follows
\bea
S_{4f} &=& {4\over \kappa^2} \int d^2 z \;
\frac{\partial^2 G(q)}
{\partial q^{ia} \partial q^{jb}}\;\psi_+^{(a} \bar \psi_+^{b)}
\chi_-^{(i}\bar \chi_-^{j)}
\label{4} \\
S_{auxf} &=& {2\over \kappa^2} \int d^2 z \;
\frac{\partial G(q)}
{\partial q^{ia}} \left( F\bar \psi_+^a \bar \chi^i_-
- \bar F \psi_+^a
\chi^i_-
+ \bar L \psi_+^a \bar \chi^i_-
+ L \bar \psi_+^a \chi^i_- \right)
\label{aux} \\
S_{kinf} &=&
{1\over \kappa^2} \int d^2 z \;\{ iG(q)\;(\partial_{++}
\chi^i_- \bar \chi_{-\;i}
-\partial_{++} \bar \chi^i_- \chi_{-\;i} +
\partial_{--} \psi^a_+ \bar \psi_{+\;a} -
\partial_{--} \bar \psi^a_+
\psi_{+\;a})
\nonumber \\
&& -2i \frac{\partial G(q)}{\partial q^{ia}} \;
(\partial_{++} q_j^{\;a}
\chi^{(j}_- \bar \chi^{i)}_-
+ \partial_{--} q^i_{\;b} \psi_+^{(b} \bar
\psi_+^{a)}) \}\;. \label{kin}
\eea

Using the explicit expressions (\ref{exprG1}), (\ref{exprG2}) for
$G(q)$, one observes:

\vspace{0.3cm}
\noindent (i). After the field redefinition
\bea
F &=& F' - 2 e^{-u} \;\tilde{q}_{ia} \;\psi^{a}_+ \;\chi^{i}_- \;,
\nonumber \\
L &=& L' + 2e^{-u}\; \tilde{q}_{ia} \;\psi^a_+ \;\bar \chi^i_-
\label{flredef}
\eea
the sum of $S_{4f}$ and $S_{auxf}$ is entirely cancelled by the
contribution coming from $S_{sc}^{bos}$. Thus the off-shell
superconformally invariant
$q^{(1,1)}$ action does not contain 4-fermionic term which in general
is present in the
generic action (\ref{genaction}). The full auxiliary fields part
of the action takes the simple form
\be
S_{sc}^{aux} = {1\over \kappa^2} \int d^2 z \;e^{-2u} (F'
\bar F' + L' \bar L')\;.
\ee

\vspace{0.3cm}
\noindent (ii). Being written through redefined fermionic fields
\be
\chi_-^a = e^{-u}\; \tilde{q}^{\;a}_i \;\chi^i_- \;,\;
\bar \chi_-^a = e^{-u}\; \tilde{q}^{\;a}_i \;\bar \chi^i_- \;, \;
\psi_+^i = e^{-u}\; \tilde{q}^{i}_{\;a}\; \psi^a_+ \;, \;
\bar \psi_+^i = e^{-u}\; \tilde{q}^{i}_{\;a}\; \bar \psi^a_+ \;,
\label{fermredef}
\ee
$S_{kinf}$ is reduced to a sum of the free fermionic terms
\be
S_{kinf} =
{1\over \kappa^2} \int d^2 z \;i \{ \partial_{++}
\chi^a_- \bar \chi_{-\;a}
-\partial_{++} \bar \chi^a_- \chi_{-\;a} +
\partial_{--} \psi^i_+ \bar \psi_{+\;i} - \partial_{--} \bar \psi^i_+
\psi_{+\;i})\;.
\label{okonkinf}
\ee

\vspace{0.3cm}
Thus, on shell $S_{sc}$ is reduced to a sum of the free action
for the scalar field $u(z)$, $SU(2)$ WZNW sigma model action
for the field $\tilde{q}^{ia}$ and the free actions for fermionic
fields $\chi_-^a$ and $\psi_+^i$. This is just the action
of $N = 4\; SU(2)\times U(1)$ WZNW sigma model
\cite{{IKL},{Belg1},{GI}}. The off-shell
component action (with the redefinitions (\ref{flredef}),
(\ref{fermredef})) is also identical to the one given in \cite{GI}.

Finally, it is worth clarifying the term ``$SU(2)\times U(1)$
WZNW sigma model'' in the
present context. The field $\tilde{q}^{ia}$ contains just three
independent
parameters-fields and so parametrizes the coset
$SU(2)_{cL}\times SU(2)_{cR}/SU(2)_{diag}$, where $SU(2)_{cL}$ and
$SU(2)_{cR}$ are rigid $SU(2)$ subgroups of two commuting left and
right light-cone branches of the full $N=4\;\; SU(2)$
superconformal group.
What is the group-theoretical meaning of the scalar field $u(z)$?
Let us remind that the entire symmetry of the considered model is
the ``large''
$N=4\; SO(4) \times U(1)$ superconformal symmetry which is a closure
of two $N=4 \; SU(2)$ ones defined in Subsect.2.4. Two local
supersymmetries present in $N=4 \;\; SU(2)$ superconformal
groups I and II contain in their
commutator $U(1)$ Kac-Moody transformations which are realized in the
given model as shifts of the field $u(z)$ by a sum of two
arbitrary holomorphic functions of, respectively,
$z^{++}$ and $z^{--}$. So, one is led to
identify $u(z)$ with a parameter of the coset
$U(1)_L\times U(1)_R /U(1)_{diag}$, and this explains the term
``$SU(2)\times U(1)$ WZNW model''. Let us stress that in the
present case the internal symmetry does not commute with
supersymmetry, in contrast, e.g., to $N=1$ supersymmetric
WZNW models \cite{Curt}. Instead,
it constitutes a nontrivial part of the underlying superconformal
symmetry.

\setcounter{equation}{0}
\section{Massive deformations of (4,4) sigma models}

A standard way to generate potential (in particular, mass)
terms in $N=4\;\; 2D$ ($N=2\;\; 4D$)
sigma model actions is to modify the supersymmetry algebra
by central charges
and to identify the latter with some isometries of the
original action \cite{{Alv},{Gates}}. As was noticed in \cite{IKL}
and, recently, in \cite{HP},
in the $(4,4)$ sigma models there exists a possibility
to add such terms without changing the supersymmetry algebra. The
explicitly elaborated example is the $N=4\;SU(2)$
WZNW - Liouville system of refs. \cite{{IK2},{IKL},{GI}} which is
a superconformally invariant deformation of the
$N=4\;\;SU(2)\times U(1)$ WZNW model discussed in the
previous Section. Here we
reproduce this example within
the harmonic superspace formalism and comment on a more
general situation when the generic action (\ref{genaction})
is subjected to a deformation of this kind. It turns out that the
deforming term in the superfield $q^{(1,1)}$ action is
defined in a unique way. As a result,
potential terms in the component action are almost completely
specified by the original target space bosonic metric.

\vspace{0.4cm}
\noindent{\bf 5.1 Deformations of the general $q^{(1,1)}$ action.}
Keeping in mind that the superfield $q^{(1,1)}$ and the integration
measure $\mu^{(-2,-2)}$ are dimensionless, the only way to
construct a manifestly analytic
massive term for $q^{(1,1)}$ is to allow for explicit
$\theta$'s in the action.

The simplest term of this kind reads
\be
S_m = m \int \mu^{(-2,-2)}\;
\theta^{(1,0)\;\underline{i}}
\theta^{(0,1)\;\underline{b}}\;
C_{\underline{i}\;\underline{b}}^M\;
q^{(1,1)\;M}
\;; \;\; [m] = cm^{-1}\;,
\label{mtermg}
\ee
where $C_{\underline{i}\;\underline{b}}^M $ are arbitrary
constants (subject to
the appropriate reality conditions), and we once again
resorted to the quartet notation (see eqs. (\ref{quartet1})).
It immediately follows that, despite the presence of
explicit $\theta$'s, (\ref{mtermg})
is invariant under rigid $(4,4)$
supersymmetry: one represents the supertranslation of, say,
$\theta^{(1,0)\;\underline{i}}$ as
$$
\delta_{SUSY}\; \theta^{(1,0)\;\underline{i}} =
\epsilon^{k\underline{i}} u^{(1)}_k
= D^{(2,0)} \epsilon^{k\underline{i}} u^{(-1)}_k\;,
$$
integrate by part with respect to $D^{(2,0)}$ and make use
of the defining constraints (\ref{constrM}).

It is easy to argue that this linear term is the
{\it only} possible supersymmetric massive term of $q^{(1,1)\;M}$.
Indeed, adding of any
higher degree monomial of $q^{(1,1)\;M}$ to (\ref{mtermg})
would require inserting explicit harmonics $u^{(-1)}_i,
\;v^{(-1)}_a$ to ensure the balance of $U(1)$ charges.
After taking off the harmonic derivatives $D^{(2,0)}, D^{(0,2)}$
from the supervariations of the analytic $\theta$'s and
integrating by part, these derivatives would hit not only
the superfields $q^{(1,1)\;M}$, but also the harmonics
$u^{(-1)}_i,\;v^{(-1)}_a$, in the latter
case with a non-vanishing result.

It is also straightforward to check that (\ref{mtermg})
respects invariance not only under rigid $(4,4)$
supersymmetry, but also under the {\it whole}
$N=4 \;\;SU(2)$ superconformal group defined by
eqs.(\ref{scgI}) - (\ref{measI}), (\ref{supquI}). At the
same time, it breaks invariance under the second
$N=4 \;\;SU(2)$ superconformal group and, hence, the
``large'' $N=4 \;\;SO(4)\times U(1)$ superconformal group.
The only additional manifest invariance one can achieve
is the diagonal $SU(2)$ in the product of two rigid $SU(2)$'s
acting on the indices $\underline{i}$ and
$\underline{b}$ of Grassmann coordinates in (\ref{mtermg}),
provided $C_{\underline{i}\;\underline{b}\;M} \sim
\epsilon_{\underline{i}\;\underline{b}}$.

Note that (\ref{mtermg}) can be rewritten in the
standard $(4,4)$ superspace as a Fayet-Iliopoulos term
of the unconstrained prepotential solving the irreducibility
conditions (\ref{twistcon}) \cite{GI}. In such
a form the mass term does not involve explicit $\theta$'s
and so is manifestly supersymmetric.

Let us examine how the adding of (\ref{mtermg}) to the
generic action (\ref{genaction}) influences the component
structure of the latter.
After integrating over
Grassmann and harmonic variables, (\ref{mtermg}) reads
\be
S_m = - 2m \int d^2 z \;
F^M_{\underline{i}\;\underline{b}}\;
C^{\underline{i}\;\underline{b}\;M}\;,
\ee
where we made use of the matrix notation (\ref{quartF})
for auxiliary fields.
After eliminating the auxiliary fields in the sum
$S_{q} + S_{m}$, the physical component action acquires
new terms: certain
Yukawa type couplings between fermions and the field $q^{ia\;M}$
as well as a potential term of the latter.
All these new terms are expressed through
the ``metric'' $G_{M\;N}(q)$ defined by eq. (\ref{metr}) and its
inverse $G^{M\;N}(q)$ ($G^{M\;N}G_{N\;K} = \delta^M_K$). We
give here explicitly only the potential term of $q^{(1,1)\;M}$
\be
S_{q}^{pot} = {{m^2}\over 2} \int d^2 z \;G^{M\;N}(q)\;
(C_{\underline{i}\;\underline{a}}^M C^{\;\underline{i}\;
\underline{a}\;N}) \;.
\ee

\vspace{0.4cm}
\noindent{\bf 5.2 $N=4 \; SU(2)$ WZNW - Liouville model.}
As an instructive
example we will discuss a massive deformation of the
superconformal action (\ref{confact}).

As was already noticed, the mass term (\ref{mtermg})
preserves the $N=4\;\;SU(2)$ superconformal symmetry I,
so the model described by the action
\bea
S_{sc}^m &=& - {1\over {4\kappa^2}} \int \mu^{(-2,-2)}
\{
\hat{q}^{(1,1)} \hat{q}^{(1,1)} \left( \frac{1}{(1+X)X} -
\frac{\mbox{ln}(1+X)}{X^2} \right) \nonumber \\
&& + 2m \;
\theta^{(1,0)\;\underline{i}}
\theta^{(0,1)\;\underline{b}}\;
C_{\underline{i}\;\underline{b}}\;
q^{(1,1)} \}
\label{wznwl}
\eea
is a superconformally invariant deformation of
$N=4\;\;SU(2)\times U(1)$ WZNW model. Actually, the mass
parameter in (\ref{wznwl}) is inessential as it
can be fixed at any non-zero value by rescaling $q^{(1,1)}$ as
$$
q^{(1,1)} \Rightarrow \gamma \; q^{(1,1)}\;.
$$
The sigma model part of the action (\ref{wznwl})
is invariant under this
rescaling because the relevant variation looks precisely
the same as the rigid dilatation one. Also, using the
invariance of the sigma model part under the group
$SO(4)\sim  SU(2)\times SU(2)$ acting on the
underlined indices (these $SU(2)$'s enter into the left and right
branches of superconformal group II) and absorbing the norm of the
four-vector $C_{\underline{i}\;\underline{b}}$ into a
renormalization of the parameter $m$, one may rotate this constant
vector into the form
$$
C_{\underline{i}\;\underline{b}} = \epsilon_{\underline{i}\;
\underline{b}}\;,
$$
which explicitly shows that (\ref{wznwl}) possesses an extra
symmetry with respect to
the diagonal $SU(2)$ from the $SO(4)$ just mentioned.

With taking account of the last remark,
in the component language the mass term in (\ref{wznwl}) reads
\be
2m \int \mu^{(-2,-2)}\;
\theta^{(1,0)\;\underline{i}}
\theta^{(0,1)\;\underline{b}}\;
\epsilon_{\underline{i}\;\underline{b}}\;
q^{(1,1)} =
4m \int d^2 z\;
F_{\underline{i}\;\underline{b}}\;
\epsilon^{\underline{i}\;\underline{b}} \;.
\ee
After eliminating auxiliary fields it gives rise to the following
physical component action of the deformed
$N=4\;\;SU(2)\times U(1)$ WZNW sigma model
\be
S_{sc(m)} = S_{sc}^{bos} (F=L=0) + S_{kinf} + S_m \;,
\label{liouv}
\ee
where $S_{sc}^{bos}$ and $S_{kinf}$ are given by eqs.
(\ref{okonb}) and (\ref{okonkinf}) and
\bea
S_m &=& {1\over \kappa^2} \int d^2 z \; \{ m^2 \;
e^{2u}
+ 2m\; e^{-u} \;\tilde{q}_{ia}\;( \bar \psi_+^a \chi^i_- -
\psi_+^a \bar \chi^i_- ) \}\;.
\label{liouvm}
\eea
The action (\ref{liouv}) is just the on-shell action of
$N=4\;SU(2)$ WZNW - Liouville
system \cite{{IK2},{IKL},{GI}}.

\setcounter{equation}{0}
\section{(4,4) duality transformation}

Up to now we dealt with the constrained analytic superfields
$q^{(1,1)}$. However, for several reasons it would be
advantageous to have unconstrained superfield formulations
of the models presented in the previous Sections. Firstly,
this seems necessary for the complete understanding of the
target space geometry hidden in the sigma model actions
(\ref{genaction}), (\ref{confact}). Secondly, such formulations
would allow to straightforwardly deduce the relevant
superfield equations of motion.
To know such equations is important, e.g.,  while analyzing
the integrability properties of given model (the existence of
a zero curvature representation,
an infinite number of superfield conserved quantities, etc.).
Thirdly, unconstrained formulations could prompt how to extend
the above formalism to accommodate more general class of $(4,4)$
sigma models with non-commuting left and right complex structures.

One way to achieve an unconstrained superfield formulation
is to express the action through prepotentials solving the harmonic
conditions
(\ref{constrM}) \cite{{Sieg},{GI}}. Unfortunately,
when doing so, one loses the privilege of manifest harmonic
analyticity as the relevant prepotential is a general $(4,4)$
superfield. Besides, it is dimensionful and therefore can hardly
be identified with any object of the target space geometry (e.g., a
coordinate on the target manifold). Therefore, like in
the $N=2\;\;4D$ case \cite{GIO2}, we prefer
another way which is based on implementing (\ref{constrM})
in the action with the help of unconstrained analytic Lagrange
multipliers. After
eliminating the original $q^{(1,1)}$ by its algebraic equation of
motion one gets a new, dual representation of the action via the
Lagrange multipliers. Below we present dual forms of the previously
constructed $q^{(1,1)}$ actions and discuss some peculiarities of
them.

\vspace{0.4cm}
\noindent{\bf 6.1 Transforming the general $q^{(1,1)}$ action.}
Let us consider the following modification of the $q^{(1,1)}$
action (\ref{genaction})
\be
S_{q,\omega} = \int \mu^{(-2,-2)} \{ {\cal L}^{(2,2)} (q^{(1,1)\;M},
u, v) + \omega^{(-1,1)\;M} D^{(2,0)} q^{(1,1)\;M} +
\omega^{(1,-1)\;M}
D^{(0,2)} q^{(1,1)\;M} \}\;.
\label{dualgen}
\ee
The analytic superfields $q^{(1,1)\;M}$, $\omega^{(1,-1)\;M}$,
$\omega^{(-1,1)\;M}$
are unconstrained and one can vary them to get the superfield
equations of motion. Varying $\omega^{(1,-1)\;M}$,
$\omega^{(-1,1)\;M}$ yields the constraints
(\ref{constrM}) and we recover the original action
(\ref{genaction}). Alternatively, one can vary (\ref{dualgen})
with respect to $q^{(1,1)\;M}$, which gives rise to the equation
\be
\frac{\partial {\cal L}^{(2,2)}}{\partial q^{(1,1)\;M}} =
D^{(2,0)} \omega^{(-1,1)\;M} + D^{(0,2)}
\omega^{(1,-1)\;M} \equiv
A^{(1,1)\;M}\;.
\label{dualeq}
\ee
This algebraic equation is a kind of Legendre transformation
allowing to express $q^{(1,1)\;M}$ as a function of
$A^{(1,1)\;M}$
\be
\mbox{(6.2)} \Rightarrow q^{(1,1)\;M} = q^{(1,1)\;M}
(A^{(1,1)}, u, v) \;.
\ee
Substituting this expression back into (\ref{dualgen}), one
arrives at the dual form of the $q^{(1,1)}$ action
\bea
S_\omega  &=& \int \mu^{(-2,-2)} {\cal L}^{(2,2)}_\omega
(A^{(1,1)}, u,v)\;,
\nonumber \\
{\cal L}^{(2,2)}_\omega (A^{(1,1)}, u,v) &\equiv &
{\cal L}^{(2,2)} (q^{(1,1)\;M} (A, u,v), u,v)
- q^{(1,1)\;M} (A,u,v)
A^{(1,1)\;M}\;.
\label{dualg2}
\eea

The dual action (\ref{dualg2}) provides a new off-shell
formulation of $(4,4)$ sigma models with commuting left and
right complex structures
via {\it unconstrained} analytic $(4,4)$ superfields. The most
characteristic
feature of such formulations is the presence of infinite number of
auxiliary fields \cite{GIKOS}. Thus, in the case at hand the
physical component action for $4n$ bosons
and $4n$ fermions is restored only after eliminating an
infinite tower of auxiliary fields coming from the double harmonic
expansion of superfields $\omega^{(1,-1)}(\zeta, u, v)\;,
\;\omega^{(-1,1)} (\zeta, u, v)$. We postpone the discussion
of the component content of the dual $(4,4)$ sigma model actions
to further publications. Here we will briefly mention only some
salient features of the dual formulation.

The action (\ref{dualg2}) has a clear analog in the
$N=2\;\;4D$ harmonic analytic superspace: it is the $\omega$
hypermultiplet action dual to the action of tensor
$N=2\;\;4D$ multiplet \cite{GIO1}. An essentially new feature
of the $(4,4)\;\;2D$ action is the presence of two
independent sets of Lagrange
multipliers  $\omega^{(1,-1)\;M},\; \omega^{(-1,1)\;M}$.
An inspection
of their field content shows that they contain twice as
many physical fields as compared with the original constrained
$q^{(1,1)}$ superfields. In particular, there are two sets of
the bosonic fields appearing as first terms in the harmonic
expansions of $\omega^{(1,-1)\;M},\; \omega^{(1,-1)\;M}$
\be
\omega^{(1,-1)\;M} (\zeta, u,v) = \omega_0^{ia\;M}
u^{(1)}_iv^{(-1)}_a
+ ... \;,
\; \omega^{(1,-1)\;M} (\zeta, u,v) =
\tilde{\omega}_0^{ia\;M}u^{(-1)}_iv^{(1)}_a + ...\;.
\ee
So one may wonder how the equivalence to the original
$q^{(1,1)}$ action is achieved at the component level.

The answer proves very simple: the actions (\ref{dualgen}) and
(\ref{dualg2}) are invariant under the gauge transformations
\be
\delta \;\omega^{(1,-1)\;M} = D^{(2,0)} \sigma^{(-1,-1)\;M}
\;, \; \delta \;\omega^{(-1,1)\;M}
= - D^{(0,2)} \sigma^{(-1,-1)\;M} \;,
\label{gauge}
\ee
with  $\sigma^{(-1,-1)\;M} = \sigma^{(-1,-1)\;M}(\zeta, u, v)$
being arbitrary analytic functions, and this
gauge freedom takes away just half of the lowest superisospin
multiplets in the superfields $\omega^{(1,-1)\;M}\;,\;
\omega^{(1,-1)\;M}$, thus restoring the correct
physical field content. For instance, the first
components of these superfields are transformed as
\be
\delta \;\omega^{(1,-1)\;M}_0 (z) = \partial^{(2,0)}
\sigma^{(-1,-1)\;M} (z)\;,\;
\delta\; \omega^{(-1,1)\;M}_0 (z) = - \partial^{(0,2)}
\sigma^{(-1,-1)\;M} (z) \;,
\ee
and one may fix the gauge so as to entirely remove one set of
these fields (other gauge choices are possible as well). Thus,
in contrast
to the $q^{(1,1)}$ superfield formulation,
where the necessary set of the physical fields is ensured by
imposing the harmonic constraints on $q^{(1,1)}$, the same
goal in the dual formulation is achieved thanks to the gauge
freedom (\ref{gauge}) (and after eliminating
an infinite set of auxiliary fields).

Note that (\ref{gauge}) is a generalization of the abelian shift
isometries of the $\omega$ hypermultiplet Lagrange multipliers
of the dual $N=2\;\;4D$ actions \cite{GIO2}. The origin of
these isometries lies in the fact that the scalar components
undergoing constant shifts are dual to the notoph field strengths
which are present in the initial
formulation via the tensor multiplet superfields, and
therefore these components always enter the action through their
$x$ derivatives.
However, such an
interpretation seems not possible for the transformations
(\ref{gauge}), because there is no analog of the notoph field
strength in the $\theta$ expansion of $q^{(1,1)}$ (\ref{compon}).
Both in the $q$ and $\omega$ languages the bosonic degrees of
freedom in the considered case are represented by the canonical
scalar fields. So it remains a mystery what is the geometric
origin of the gauge transformations (\ref{gauge}) (see, however,
the next comment).

One more interesting feature of the dual formulation is
revealed while applying the $(4,4)$ duality transformation
to the mass term deformed action
$$
S_q + S_m \;,
$$
where $S_m$ is given by eq. (\ref{mtermg}). The algebraic equation
(\ref{dualeq}) is now modified by $\theta$ dependent terms
\be
\frac{\partial {\cal L}^{(2,2)}}{\partial q^{(1,1)\;M}} =
A^{(1,1)\;M}
-
m \theta^{(1,0)\;\underline{i}}
\theta^{(0,1)\;\underline{b}}\;
C_{\underline{i}\;\underline{b}}^M \;.
\label{dualeq1}
\ee
Accordingly, $\theta$ terms appear in the corresponding
dual action. It is seen from (\ref{dualeq1}) (or directly
from considering
(\ref{dualgen}) with the mass term added) that in this
dual formulation $(4,4)\;\;2D$ supersymmetry is realized in
a non-standard way
\bea
\delta_{SUSY}\; \omega^{(1,-1)\;M}
&=&
-m \;
\epsilon_R^{a\underline{b}}\;
v^{(-1)}_a \;
\theta^{(1,0)\;\underline{i}} \;C_{\underline{i} \;
\underline{b}}^M \nonumber \\
\delta_{SUSY}\; \omega^{(-1,1)\;M}
&=&
m \;\epsilon_L^{i\underline{i}}\; u^{(-1)}_i \;\theta^{(0,1)\;
\underline{b}} \;
C_{\underline{i} \;
\underline{b}}^M \;.
\label{susycentr}
\eea
Here $\epsilon_L^{i\underline{i}}, \;\epsilon_
R^{a\underline{b}}$ are the
constant parameters of the left and right supertranslations
in the quartet notation.
To reveal the meaning of this modification, let us consider
the Lie bracket of the left and right supertranslations.
While in the original $q$
formulation they commute irrespective of the presence or
absence of the mass
term in the action, in the realization on the superfields
$\omega$ their bracket surprisingly turns out non-vanishing.
Namely, in the obvious notation,
\be
[\delta_L, \delta_R] \omega^{(1,-1)} =
m \;\epsilon_L^{i\underline{i}}\; \epsilon_R^{b\underline{b}}\;
\;C_{\underline{i} \;
\underline{b}}^M u^{(1)}_i v^{(-1)}_b\;,\;\;
[\delta_L, \delta_R] \omega^{(-1,1)} =
- m\; \epsilon_L^{i\underline{i}}\; \epsilon_R^{b\underline{b}}\;
C_{\underline{i} \; \underline{b}}^M u^{(-1)}_i v^{(1)}_b\;.
\ee
But this is just a subclass of transformations (\ref{gauge})
with $z$ independent parameters homogeneous in harmonics
\be
\sigma^{(-1,-1)\;M}_0 = m\; \sigma_0^{i\underline{i}\;
b\underline{b}}
\;C_{\underline{i} \; \underline{b}}^M\; u^{(-1)}_i\; v^{(-1)}_b\;.
\label{centrzar}
\ee
Thus the realization (\ref{susycentr}) corresponds to the following
extension of the standard $2D$ $(4,4)$ SUSY algebra
\be
\{ Q_+^{i\underline{i}}, Q_-^{a\underline{a}} \} \sim m
Z^{i\underline{i} \;a \underline{a}}\;,
\label{ccmod}
\ee
where the ``semi-central'' charge generator
$Z^{i\underline{i} \;a \underline{a}}$ is realized by shifts
(\ref{gauge}), (\ref{centrzar}), with
$\sigma_0^{i\underline{i}\; b\underline{b}}$ being the associated
transformation parameters.

Note that the effect of activating central charge operators in
the SUSY algebras
by duality transformations is well known (see, e.g., ref.
\cite{SYam} and, in context of $N=2\;\;4D$ harmonic superspace,
ref. \cite{GIO2}). For instance,
one may introduce a mass term similar to (\ref{mtermg}) in the
general
analytic superfield action of tensor $N=2\;\;4D$ multiplets
\cite{GIO2} with preserving the original realization of $4D$
supersymmetry. After passing to the
dual $\omega$ hypermultiplet action one then finds that on
$\omega$ superfields the supersymmetry algebra is realized
with a non-zero central charge proportional to an abelian shift
isometry generator. An unusual feature of the case under
consideration is that the operator $Z^{i\underline{i} \;a
\underline{a}}$ transforms in general as a non-trivial tensor of the
automorphism group (hence the term ``semi-central''), while
in the $N=2\;\;4D$ example just mentioned a similar operator is
a pure singlet of the automorphism $SU(2)$.
Note that the appearance of ``quaternionic'' central charges in
certain $(4,4)$ supersymmetric models with non-zero scalar fields
potentials has been already observed in \cite{Towns}.

The relationship with central charges could clarify the
geometric meaning of the
strange gauge invariance (\ref{gauge}): perhaps, it can be viewed
as a gauging of the transformations (\ref{centrzar}).

In the rest of this Section we will present the dual actions for
$N=4 \;\;SU(2)\times U(1)$ WZNW sigma model and its
Liouville extension.

\vspace{0.4cm}
\noindent{\bf 6.2 Dual form of N=4 WZNW and WZNW - Liouville
actions.}
The extended action (\ref{dualgen}) specialized to
$N=4\;\;SU(2)\times U(1)$ WZNW model can be chosen as
\be
S_{sc\;(q,\omega)} = -{1\over 4\kappa^2} \int \mu^{(-2,-2)}\; \{
{\cal L}^{(2,2)}_{sc} + {1\over 2} \omega^{(-1,1)} D^{(2,0)}
q^{(1,1)} +
{1\over 2} \omega^{(1,-1)}
D^{(0,2)} q^{(1,1)} \}\;.
\label{dualsc}
\ee
By varying (\ref{dualgen}) with respect to the involved superfields
we obtain the equations of motion of $N=4\;\;SU(2)\times U(1)$
WZNW model in the analytic superspace
\be
q^{(1,1)} \frac{1}{(1+X)^2} = - {1\over 2} A^{(1,1)}\;,\;\;
D^{(2,0)}q^{(1,1)} =
D^{(0,2)} q^{(1,1)} = 0 \;.
\label{eqnmo}
\ee
First of them is algebraic and it serves to express $q^{(1,1)}$ via
$A^{(1,1)}$
\be
q^{(1,1)} = -2 \frac{A^{(1,1)}}{(1 + \sqrt{1+2A})^2}\;,\;\;\;
A \equiv c^{(-1,-1)} A^{(1,1)}\;.
\label{qA}
\ee
The second pair of equations in (\ref{eqnmo}), being kinematical
constraints in the original formulation, becomes the dynamical
equations in the dual description; substitution of the expression
(\ref{qA}) into them yields a closed set of the equations of
motion for the superfields $\omega^{(1,-1)}, \omega^{(-1,1)}$.
They can be idependently deduced
from the action obtained by substituting (\ref{qA}) into
(\ref{dualsc})
\be
S_{sc\;(w)} =
-{1\over 4\kappa^2} \int \mu^{(-2,-2)}
\left( \frac{\tilde{A}^{(1,1)}}{\tilde{A}} \right)^2
\left( \tilde{A}^2 - \tilde{A} + \mbox{ln}(1+\tilde{A}) \right)\;,
\label{dualA}
\ee
with
\be
\tilde{A}^{(1,1)} = \frac{A^{(1,1)}}{1 + \sqrt{1 + 2A}}\;, \;\;
\tilde{A} \equiv c^{(-1,-1)} \tilde{A}^{(1,1)}\;.
\ee
The action (\ref{dualA}) provides a dual description of
$N=4\;\;SU(2)\times U(1)$ WZNW sigma model in terms of
unconstrained analytic $(4,4)$ superfields $\omega^{(1,-1)},
\omega^{(-1,1)}$.
Recall that they are subjected to the gauge freedom (\ref{gauge}),
which ensures the correct physical fields content of the action.

With making use of the explicit form of the superconformal
variation of ${\cal L}_{sc}^{(2,2)}$ (\ref{var}), it is easy
to see that the action
(\ref{dualsc}) is still invariant under the transformations
(\ref{transfhat}), provided Lagrange multipliers transform as
\bea
\delta_I \;\omega^{(1,-1)} &=& -\Lambda^{(0,0)}_I \left(
\omega^{(1,-1)} +
2 c^{(1,-1)} \frac{1}{(1+X)^3} \right)\;,   \nonumber \\
\delta_I \;\omega^{(-1,1)} &=& -\Lambda^{(0,0)}_I \left(
\omega^{(-1,1)} -
2c^{(-1,1)} \frac{1-X}{(1 + X)^3} \right)\;.
\label{scomega}
\eea
Transformations from the right light-cone branch are given
by the same formulas, with the $U(1)$ charges interchanged as
$1 \leftrightarrow -1$. Substituting in (\ref{scomega}) the
expression for $q^{(1,1)}$ (\ref{qA}), one may rewrite
(\ref{scomega}) entirely in terms of
the $\omega$ fields. The resulting transformations realize the
superconformal group I in the dual $\omega$ formulation of $N=4\;\;
SU(2)\times U(1)$ WZNW model. Of course, the above actions
are invariant under the superconformal group II, with respect
to which  $\omega^{(1,-1)}, \omega^{(-1,1)}$ are scalars
like $q^{(1,1)}$.

Finally, let us obtain the dual form of WZNW - Liouville action
(\ref{wznwl}). The relevant extended action is given by a sum
of (\ref{wznwl}) and the same Lagrange multipliers term as in
(\ref{dualsc}). Then, passing to the dual formulation is
accomplished by the following
replacement in the above equations
\be
A^{(1,1)} \Rightarrow A^{(1,1)} - 4m(
\theta^{(1,0)} \bar \theta^{(0,1)} +
\theta^{(0,1)} \bar{\theta}^{(1,0)})\;.
\label{replace}
\ee
The modified expression for $q^{(1,1)}$ reads
\bea
\hat{q}^{(1,1)} &=& -2 \frac{A^{(1,1)}}{(1 + \sqrt{1+2A})^2}
+ 4m \frac{1}{\sqrt{1+2A}}
(\theta^{(1,0)} \bar \theta^{(0,1)} +
\theta^{(0,1)} \bar{\theta}^{(1,0)}) \nonumber \\
&& - 16m^2 \frac{c^{(-1,-1)}}{(1+2A)^{3/2}}
\;
\theta^{(1,0)} \bar \theta^{(1,0)} \theta^{(0,1)}
\bar \theta^{(0,1)} \;.
\eea
Accordingly, the dual
$N=4$ WZNW - Liouville action is given by
\bea
S_{(sc)m}^{dual} &=&
-{1\over 4\kappa^2} \int \mu^{(-2,-2)} \{
\left( \frac{\tilde{A}^{(1,1)}}{\tilde{A}} \right)^2
\left( \tilde{A}^2 - \tilde{A} + \mbox{ln}(1+\tilde{A}) \right)
\nonumber \\
&& + 2m \frac{1}{1+\tilde{A}}\; \tilde{A}^{(1,1)}\; (
\theta^{(1,0)} \bar \theta^{(0,1)} +
\theta^{(0,1)} \bar{\theta}^{(1,0)}) \nonumber \\
&& - 8 m^2 \frac{1}{(1+2\tilde{A})
(1+\tilde{A})}\;
\theta^{(1,0)} \bar \theta^{(1,0)} \theta^{(0,1)}
\bar \theta^{(0,1)} \}\;.
\eea
This action is invariant under the appropriate semi-central
charge modification of superconformal transformations
(\ref{scomega})
\be
\delta_I^{mod} \omega^{(1,-1)} = \delta_I \omega^{(1,-1)}\;,\;
\delta_I^{mod} \omega^{(-1,1)} = \delta_I \omega^{(-1,1)} +
im (\frac{\partial a}{\partial \bar \theta^{(1,0)}} \bar
\theta^{(0,1)} +
\frac{\partial a}{\partial \theta^{(1,0)}} \theta^{(0,1)})
\ee
(and analogously for transformations from the right light-cone
branch). Note in this connection that the substitution
(\ref{replace}) can be interpreted as the following central-charge
modification
of the harmonic derivatives  $D^{(2,0)},\;D^{(0,2)}$ in the
$\omega$ superfield strength
$A^{(1,1)}$
\bea
D^{(2,0)}
&\Rightarrow & D^{(2,0)} - 2m
(\theta^{(1,0)} \bar \theta^{(0,1)} +
\theta^{(0,1)} \bar{\theta}^{(1,0)}) Z^{(1,-1)} \nonumber \\
D^{(0,2)}
&\Rightarrow & D^{(0,2)} + 2m
(\theta^{(1,0)} \bar \theta^{(0,1)} +
\theta^{(0,1)} \bar{\theta}^{(1,0)}) Z^{(-1,1)} \;,
\eea
where the generators  $Z^{(1,-1)},\; Z^{(-1,1)}$ act on the
$\omega$
superfields as shifts
\be
Z^{(1,-1)} \omega^{(-1,1)} = 1\;,\;\;\;
Z^{(-1,1)} \omega^{(-1,1)} = -1\;,
\ee
and are the appropriate harmonic projections of the operator
appearing in (\ref{ccmod})
$$
Z^{(\pm 1, \mp 1)} \sim Z^{ia \underline{i}
\underline{a}}\epsilon_
{\underline{i} \underline{a}} u^{(\pm 1)}_i v^{(\mp 1)}_a \;.
$$

We end with two comments.

First of all, we stress that the $(4,4)$ duality transformation
discussed in this Section is a natural generalization of the
analogous transformation
proposed in \cite{GIO2} in the framework of
$N=2\;\;4D$ harmonic superspace.

The latter transformation has been used to give a constructive
proof of the statement that all self-interactions of the
constrained matter $N=2$ multiplets are
equivalent to some particular classes of the general
self-interaction of the universal unconstrained $N=2$ multiplet,
the analytic $q^+_i$ hypermultiplet
\cite{{GIKOS},{GIO2},{GIO1}}. Representing $q^+_i$ as a pair of
analytic superfields,
$q^+_i \propto \{ L^{++},\;\omega \}$, and
eliminating $L^{++}$ by its algebraic equation of motion,
the general $q^+$ action can also be written as
a general action of self-interacting $\omega$ hypermultiplets
\cite{GIOSgeo}.
These general actions are characterized by lacking of any
isometries,
while this is not the case for their particular cases
corresponding to the constrained matter $N=2$ multiplets.

In our case the direct analog of the $N=2 \;\;\omega$
hypermultiplet is the pair of unconstrained analytic
superfields $\omega^{(1,-1)},\;\omega^{(-1,1)}$ and one of
the chracteristic features of the dual action (\ref{dualg2})
is the gauge invariance (\ref{gauge}) which serves to remove
the doubling of physical degrees of freedom. This
invariance substitutes the isometries of the dual
$N=2\;\;4D$ $\omega$ hypermultiplet actions. Then, by analogy, one
might conjecture that the most general
class of $(4,4)\;\;2D$ sigma models (including those with
non-commuting left and right complex structures) is described
by the general $\omega^{(1,-1)},\;\omega^{(-1,1)}$ action
\footnote{It seems that in the
present case the notion of $q^+$ hypermultiplet is not so
useful as in the $N=2\;\;4D$ case because in order to accommodate,
e.g., the triple of superfields $q^{(1,1)},\;\omega^{(1,-1)},\;
\omega^{(-1,1)}$
(the $(4,4)$ analog of the pair $L^{++},\; \omega$) one needs two
such hypermultiplets related by some algebraic constraint.}
respecting no gauge
invariance (\ref{gauge}).
Such action definitely cannot be reformulated
in terms of constrained $q^{(1,1)}$ superfields and
so cannot be written through pairs of chiral and twisted chiral
$(2,2)$ superfields. An alternative, perhaps more attractive option
could be to somehow non-abelize the
gauge freedom (\ref{gauge}), still avoiding the doubling of
physical degrees of freedom. We hope to analyze these possibilities
in further publications.

Second remark concerns the relationship with the duality
transformation elaborated in ref.\cite{{RSS},{RASS}} for
$N=4 \;\;SU(2)\times U(1)$ WZNW model
in the formulation through chiral and twisted chiral
$(2,2)$ superfields.
This kind of duality transformation replaces the twisted multiplet
by a chiral one and so brings the model
into the torsionless form of $(2,2)$ K\"ahler sigma model for two
chiral superfields with a specific K\"ahler potential.
Though we do not know as yet
the precise relation of this transformation to ours,
the basic difference between either seems to lie in the fact
that the former breaks manifest $(4,4)$ supersymmetry while the
latter respects it at any stage. We postpone the full analysis of
the component structure of the dual $N=4\;\;SU(2)\times U(1)$
action (\ref{dualA}) to the future. However, some
preliminary tests show that after fixing a proper
gauge with respect to (\ref{gauge}) and eliminating an infinite
tower of auxiliary fields, the action of physical bosonic fields
proves the same as in the original $q^{(1,1)}$ formulation,
thus indicating that the geometry of bosonic manifold is not
affected by the $(4,4)$ duality transformation in
contrast to the aforementioned $(2,2)$ one. This point deserves
a further study.

\setcounter{equation}{0}
\section{Conclusions}

In this paper we have introduced the basic concepts of $(4,4)\;\;2D$
harmonic superspace with two independent sets of $SU(2)$ harmonics
in the left and right light-cone sectors, and constructed in its
framework off-shell superfield
actions which describe $(4,4)$ supersymmetric sigma models
with commuting left and right complex structures and provide massive
deformations of these models.
We discussed both the case of
general bosonic target manifolds of this kind and the special case
of the $SU(2)\times U(1)$ group manifold WZNW sigma model.
The analytic superfield action of the latter has been shown to
unambiguously follow from the
requirement of $N=4\;\;SU(2)$ superconformal invariance,
quite similarly to the action of improved tensor $N=2$ multiplet
in harmonic $N=2\;\;4D$ superspace. Besides the formulation in
terms of constrained analytic $q^{(1,1)}$ superfields which is
basically equivalent to the formulations via constrained
superfields in the projective \cite{{GHR},{RSS}} or
conventional \cite{GI} $(4,4)$ superspaces, we have found new
formulations of these models via unconstrained analytic
superfields $\omega^{(1,-1)}, \omega^{(-1,1)}$ with infinite
sets of auxiliary fields and
with a gauge invariance which serves to remove the doubling of
propagating degrees of freedom.
We achieved this with the help of $(4,4)$ duality transformation
which directly generalizes the duality transformation defined
earlier for sigma models
in harmonic $N=2\;\; 4D$ superspace. Some interesting features of
the dual actions have been found, in particular the appearance,
after passing to the dual formulation of massive $q^{(1,1)}$
models, of the $SU(2)$ tensor ``semi-central'' charges in the
anticommutators of the left and right $2D$ super Poincar\'e
generators.

We conclude by listing some further problems we hope to solve with
the help of the $(4,4)$ harmonic superspace formalism. \\

\noindent(A). {\it Constructing off-shell
formulations
of $(4,4)$ sigma models with non-commuting left and right
complex structures.} Some conceivable ways of approaching this
difficult problem within the $(4,4)$ harmonic superspace were
already indicated in the end of previous Section. The dual
formulations seem especially promising in this respect. \\

\noindent(B). {\it Extending the integrability concepts to the
harmonic superspace.} In Sect. 5 we have shown the uniqueness of
massive deformation of general $q^{(1,1)}$ action (without
modifying $(4,4)$ supersymmetry): the relevant component
potential terms are almost entirely (up to a constant matrix)
specified by the metric of
the original bosonic manifold. This effect is quite new, e.g.,
in comparison with the $(2,2)$ models where the superpotential
terms can be introduced independently of the sigma model term as
arbitrary holomorphic functions of chiral (or
twisted chiral) superfields. The  $N=4\;\; SU(2)\times U(1)$
WZNW model generates in this way the $N=4$ WZNW - Liouville system.
The superfield equations of the latter (as well as those of the
initial sigma model), while written in conventional $(4,4)$
superspace, are integrable in the sense that they amount to the
vanishing of some supercurvature \cite{IK2}. So, this system
has actually proved the first example of integrable
$(4,4)$ supersymmetric model. It is interesting to see how the same
integrability properties manifest themselves  when the $N=4$ WZNW -
Liouville system is represented by the equations (\ref{eqnmo}) in the
analytic harmonic superspace. Do these equations admit a
zero-curvature interpretation? To answer this question seems
especially important, having in perspective to construct
$(4,4)$ extensions of the Toda systems, both conformal and affine
(including the sine-Gordon model). We conjecture that these
systems are described by the deformed actions of the type
$S_q + S_m$, with $S_m$ defined in (\ref{mtermg})
and appropriately chosen metric functions
$G(q)_{M\;N}$. It is highly desirable to have convenient manifestly
supersymmetric criterions of integrability of the corresponding
analytic superfield equations of motion. \\

\noindent(C). {\it Coupling to $(4,4)$ world-sheet
conformal supergravity.}
To know the $(4,4)$ sigma models - supergravity couplings is
important for constructing self-consistent superstring and
higher $p$ branes models with the sigma model target manifolds
as a background. We hope that the
conformal supergravity in the analytic $(4,4)\;\;2D$
superspace can be constructed by analogy with that in the
$N=2\;\;4D$ analytic superspace \cite{GIOSgrav}. On the other hand,
it has been argued in \cite{{Rostand},{GI},{Troost}} that,
e.g., the component action of $N=4 \;\;SU(2)$ WZNW - Liouville
model itself can be interpreted as a result of fixing appropriate
gauges in the action of the Polyakov type $N=4\; 2D$ supergravity.
It would be interesting to understand how this occurs in the
framework of the analytic superfield description.

\setcounter{equation}{0}
\def\thesection{ }
\noindent\section{Acknowledgements}
We thank F. Delduc, S. Krivonos, V. Ogievetsky and, especially,
E. Sokatchev for interest in the work and illuminating discussions.
E.I. thanks Physikalisches Institut in Bonn for hospitality.
This investigation has been supported in part by the Russian
Foundation of Fundamental Research, grant 93-02-3821.

\vspace{1cm}
{\Large\bf Appendix}
\def\theequation{A.\arabic{equation}}

\vspace{0.4cm}
\noindent Here we compute the double harmonic integral
\be
G(q) = \int du dv \; \frac{1-X}{(1+X)^3}\;, \;\;
X = c^{(-1,-1)}\hat{q}^{ia}(z)
u^{(1)}_iv^{(1)}_a\;,
\label{harmbas}
\ee
which is the basic object of the
component $N=4\;\; SU(2)\times U(1)$ WZNW
action.

To simplify our task,  we will closely follow ref. \cite{GIO1}
where a heave use of different $SU(2)$'s realized on the fields
and harmonics has been made to compute analogous integrals.

First of all, we exploit the invariance of the integration
measure in (\ref{harmbas}) under two independent $SU(2)$
rotations of harmonics in indices $i$ and $a$. Using this
freedom, one may bring $c^{ia}$ into the form
\be
c^{ia} = \epsilon^{ia}\;.
\label{chooC}
\ee

Secondly, it is straightforward to check that (\ref{harmbas})
is invariant under the transformations of rigid $SU(2)$
subgroups of the left and right branches of $N=4\;\;SU(2)$
superconformal group I. Recall that these transformations,
e.g. from the left light-cone sector, are given by
\bea
\delta_{SU(2)_c}\; \hat{q}^{(1,1)}_0 &=& \lambda^{(ij)}u^{(1)}_i
u^{(-1)}_j
\;(\hat{q}^{(1,1)} + c^{(1,1)}) - \lambda^{(ij)} u^{(1)}_i
u^{(1)}_j \;
c^{(-1,1)} \nonumber \\
\delta_{SU(2)_c}\; u^{(1)}_i &=&
\lambda^{(kj)} u^{(1)}_k u^{(1)}_j\;u^{(-1)}_i
\;,\;\;\; \delta_{SU(2)_c}\; u^{(-1)}_i = 0\;.
\eea
Under these transformations and their right counterparts the
field  $q^{ia}(z) = \hat{q}^{ia} + c^{ia}$ undergoes independent
$SU(2)$ rotations in indices $i$ and $a$, so one may choose
the frame where
\be
q^{ia} = \epsilon^{ia}\; \rho (z)\;, \;\; \rho^2 =
{1\over 2} \;q^{ia}q_{ia}\;.
\label{chooQ}
\ee
With the choice (\ref{chooC}), (\ref{chooQ}):
\be
X = (\rho -1)\; (u^{(-1)}_2 v^{(-1)}_1 - u^{(-1)}_1 v^{(-1)}_2 )
(u^{(1)}_2 v^{(1)}_1 - u^{(1)}_1 v^{(1)}_2) \;.
\label{chooX}
\ee

Next simplifying step is to represent the integrand in
(\ref{harmbas}) as
\be
\frac{1-X}{(1+X)^3} = \left( 1 + 3\frac{\partial}{\partial \alpha} +
\alpha \frac{\partial^2}{\partial \alpha^2} \right)
\frac{1}{1+\alpha X}
\;|_{\alpha = 1}\;,
\label{alpha}
\ee
thereby reducing the problem to the
computation of the harmonic
integral
\be
I_\alpha (q) \equiv \int du dv \;\frac{1}{1+\alpha X}\;.
\label{intI}
\ee
Choosing the Euler angle parametrization for the harmonics
\bea
u^{(1)}_1 &=& i\; \mbox{sin}\; \theta /2 \;e^{-i\phi/2}\;,\;
\;u^{(-1)}_1 \;=\; \mbox{cos}
\; \theta/2\;e^{-i\phi/2}\;, \nonumber \\
u^{(1)}_2 &=& \mbox{cos}\; \theta /2 \;e^{i\phi/2}\;,\; \;\;\;\;\;
u^{(-1)}_2 \;=\;
i\; \mbox{sin}
\;\theta/2\;e^{i\phi/2}\;, \nonumber \\
v^{(1)}_1 &=& i\; \mbox{sin}\; \omega /2 \;e^{-i\gamma/2}\;,\;\;
v^{(-1)}_1 \;=\;
\mbox{cos}
\;\omega/2\;e^{-i\gamma/2}\;, \nonumber \\
v^{(1)}_2 &=& \mbox{cos}\; \omega /2 \;e^{i\gamma/2}\;,\;\;\;\;\;
v^{(-1)}_2 \;=\;
i\; \mbox{sin}
\;\omega/2\;e^{i\gamma/2}\;,
\label{param}
\eea
one rewrites (\ref{intI})
as
\bea
I_\alpha (q) &=& \int du dv\; \{ 1 + {\alpha\over 2}(\rho -1)
[ 1-\mbox{cos}\; \theta \;\mbox{cos}\; \omega - \mbox{sin}\;
\theta \;\mbox{sin}\; \omega
\;\mbox{cos}\; (\phi - \gamma)] \}^{-1}\;, \nonumber \\
\int du dv &=& {1\over {(4\pi)^2}} \int^\pi_0 \mbox{sin}\;
\theta\; d\theta
\int^\pi_0 \mbox{sin}\; \omega\; d\omega \int^{2\pi}_0 d\phi
\int^{2\pi}_0
d\gamma \;.
\eea
A straightforward calculation yields
\be
I_\alpha (q) = \frac{\mbox{ln} (1 + \alpha (\rho -1))}
{\alpha (\rho -1)}\;,
\ee
whence, using (\ref{alpha}),
\be
G(q) = \rho^{-2}\;.
\ee

Note that (\ref{intI}) could be computed without resorting to the
explicit parametrization of harmonics, by expanding (\ref{intI})
in a series in $\alpha$ and applying the formal rules of
integration over harmonic variables (\ref{int}).

Finally, we notice that the above method is not directly
applicable for computing the
harmonic integrals in the torsion potential (\ref{torssc}). The
latter is not a tensor under the action of groups $SU(2)_{cL}$,
$SU(2)_{cR}$: it is shifted by full $z$ derivatives and
its $SU(2)$ variations vanish only after
performing $z$ integration. As a result, one cannot choose $q^{ia}$
in the simple form (\ref{chooQ}), while performing the
$u,v$ integration. As we know, this difficulty can be got over by
passing to the torsion
field strength $H_{ia\; jb\; kd}$ which have tensor properties
with respect to superconformal $SU(2)$'s.

\setcounter{equation}{0}

\end{document}